\documentclass[12pt,preprint]{aastex}
\def\gapprox{\;\rlap{\lower 2.5pt                       % approx. greater
             \hbox{$\sim$}}\raise 1.5pt\hbox{$>$}\;}

\shorttitle{The Homogeneity of Interstellar Oxygen}
\shortauthors{Cartledge, Meyer, Lauroesch \& Sofia}

\begin{document}

\title{The Homogeneity of Interstellar Oxygen in the Galactic
Disk\footnote{Based on observations with the NASA/ESA \textit{Hubble Space
Telescope (HST)} and the NASA-CNES-CSA \textit{Far-Ultraviolet Spectroscopic
Explorer (FUSE)}. \textit{HST} spectra were obtained at the Space Telescope
Science Institute, which is operated by the Association of Universities for
Research in Astronomy, Inc. under NASA contract No. NAS 5-26555; \textit{FUSE}
is operated for NASA by the Johns Hopkins University under NASA contract
NAS-32985.}}

\author{Stefan I. B. Cartledge}
\affil{Department of Physics and Astronomy, Louisiana State University,
    Baton Rouge, LA 70803}
\email{scartled@lsu.edu}

\author{J. T. Lauroesch, David M. Meyer}
\affil{Department of Physics and Astronomy, Northwestern University,
    Evanston, IL 60208}
\email{jtl@elvis.astro.nwu.edu, davemeyer@northwestern.edu}

\and

\author{U.J. Sofia}
\affil{Department of Astronomy, Whitman College, Walla Walla, WA 99362}
\email{sofiauj@whitman.edu}

\begin{abstract}
We present an analysis of high resolution \textit{HST} Space Telescope Imaging
Spectrograph (STIS) observations of \ion{O}{1} $\lambda$1356 and \ion{H}{1}
Lyman-$\alpha$ absorption in 36 sight lines that probe a variety of Galactic
disk environments and include paths that range over nearly 4 orders of magnitude
in $f$(H$_2$), over 2 orders of magnitude in ${\langle}n_{\rm H}\rangle$, and
that extend up to 6.5 kpc in length. Since the majority of these sight lines
have also been observed by the \textit{Far-Ultraviolet Spectroscopic Explorer
(FUSE)}, we have undertaken the study of gas-phase O/H abundance ratio
homogeneity using the current sample and previously published Goddard
High-Resolution Spectrograph (GHRS) results. Two distinct trends are identified
in the 56 sight line sample: an apparent decrease in gas-phase oxygen abundance
with increasing mean sight line density (${\langle}n_{\rm H}\rangle$) and a gap
between the mean O/H ratio for sight lines shorter and longer than about 800 pc.
The first effect is a smooth transition between two depletion levels associated
with large mean density intervals; it is centered near ${\langle}n_{\rm
H}\rangle = 1.5 {\rm cm}^{-3}$ and is similar to trends evident in gas-phase
abundances of other elements. Paths less dense than the central value exhibit a
mean O/H ratio of log$_{10}$(O/H) = -3.41$\pm$0.01 (or 390$\pm$10 $ppm$), which
is consistent with averages determined for several long, low-density paths
observed by STIS \citep{and03} and short low-density paths observed by
\textit{FUSE} \citep{moo02}. Sight lines of higher mean density exhibit an
average O/H value of log$_{10}$(O/H) = -3.55$\pm$0.02 (284$\pm$12 $ppm$). The
datapoints for low-${\langle}n_{\rm H}\rangle$ paths are scattered more widely
than those for denser sight lines, due to O/H ratios for paths shorter than 800
pc that are generally about 0.10 dex lower than the values for longer ones.
Scenarios which would be consistent with these results include a recent infall
of metal-poor gas onto the local Galactic disk and an interstellar environment
toward Orion that is conducive to reducing the apparent gas-phase oxygen
abundance.
\end{abstract}


\section{Introduction}
Ranking as the third most abundant element in the interstellar medium (ISM)
after hydrogen and helium, oxygen is a primary constituent of both its gas and
dust phases. Consequently, accurate measurements of interstellar oxygen
abundances and the homogeneity of its distribution are crucial to a broad range
of fields, from galactic chemical evolution \citep{tim95} to the composition of
interstellar dust grains \citep{sof01}. Efforts to determine the
interstellar oxygen abundance and investigate its homogeneity using Goddard
High-Resolution Spectrograph (GHRS) observations of the optically-thin
1356{\AA} resonance line initially showed that within about 500 pc of the Sun,
the gas-phase O/H ratio for 13 sight lines was remarkably uniform at a level of
log$_{10}$(O/H) = $-$3.46$\pm$0.02 ($343\pm15$ $ppm$, \citealt{mey98}; the
original result has been linearly adjusted to reflect the $f$-value used in
this paper). It was concluded that the apparent scatter in the local diffuse
ISM O/H gas-phase abundance ratio was consistent with the sum of observational
and measurement uncertainties, and that the total oxygen abundance and the
proportions in gas and dust did not vary significantly in the diffuse ISM. More
recent studies based on spectral data from Space Telescope Imaging Spectrograph
(STIS) and \textit{Far-Ultraviolet Spectroscopic Explorer (FUSE)}, however,
have challenged this picture of homogeneity by identifying sight lines with a
much larger range of gas-phase O/H ratios. For instance, \citet{car01} reported
the first indications of enhanced oxygen depletion when they examined several
sight lines with large mean hydrogen sight line densities (${\langle}n_{\rm
H}\rangle$), \citet{moo02} measured an O/H ratio of log$_{10}$(O/H) =
$-$3.52$\pm$0.03 (303$\pm$21 $ppm$) toward 5 dwarf stars within 200 pc, and
\citet{and03} determined a mean of log$_{10}$(O/H) = $-$3.39$\pm$0.01
(408$\pm$13 $ppm$) for 19 low mean density sight lines with lengths of 0.84 --
5.01 kpc.

The distinctions between these measurements appear to be related to properties
of the samples observed by each of the different instruments. In examining the
variation of deuterium to oxygen abundance ratios in the local ISM using
\textit{FUSE}, \citet{ste03} concluded that one of the sight lines studied by
\citet{moo02} may have biased their sample unfairly in terms of the weighted
mean. Consequently, the O/H abundance ratio from \textit{FUSE} data might be
thought of as log$_{10}$(O/H) = $-$3.41$^{+0.03}_{-0.04}$ ($393\pm$32 $ppm$;
\citealt{moo02,ste03}) with one strongly deviating sight line. The revised mean
is then consistent with the long sight line STIS average of \citet{and03}, but
these values remain roughly 0.10 dex larger than the GHRS results for similar
sight lines. In an effort to develop a more thorough understanding of
variations in the interstellar oxygen abundance, and possibly explain the
differences evident in the samples studied to date, we present \ion{O}{1}
$\lambda$1356 and \ion{H}{1} $\lambda$1216 data for 36 sight lines and analyze
variations in O/H abundance ratios for 56 paths toward stellar objects observed
by GHRS, STIS, and \textit{FUSE}. The current data include sight lines
exhibiting a broad range of properties [e.g., log$_{10}{\langle}n_{\rm
H}\rangle = -1.15$ -- 1.10 (cm$^{-3}$), log$_{10}f($H$_2) = -5.22$ -- $-0.31$,
$E(\bv) = 0.03$ -- 0.66, $d = 0.15$ -- 6.56 kpc], and suggest that the
gas-phase O/H abundance ratio can be expressed as a function of
${\langle}n_{\rm H}\rangle$ where oxygen depletion is significantly enhanced
for sight lines with mean densities larger than about 1.5 cm$^{-3}$.
Furthermore, it is apparent that among the less dense paths, those extending
farther than 800 pc possess larger oxygen abundances than shorter ones.

\section{Observations and Data Extraction} 
\label{section_observations}
\begin{figure}
\plotone{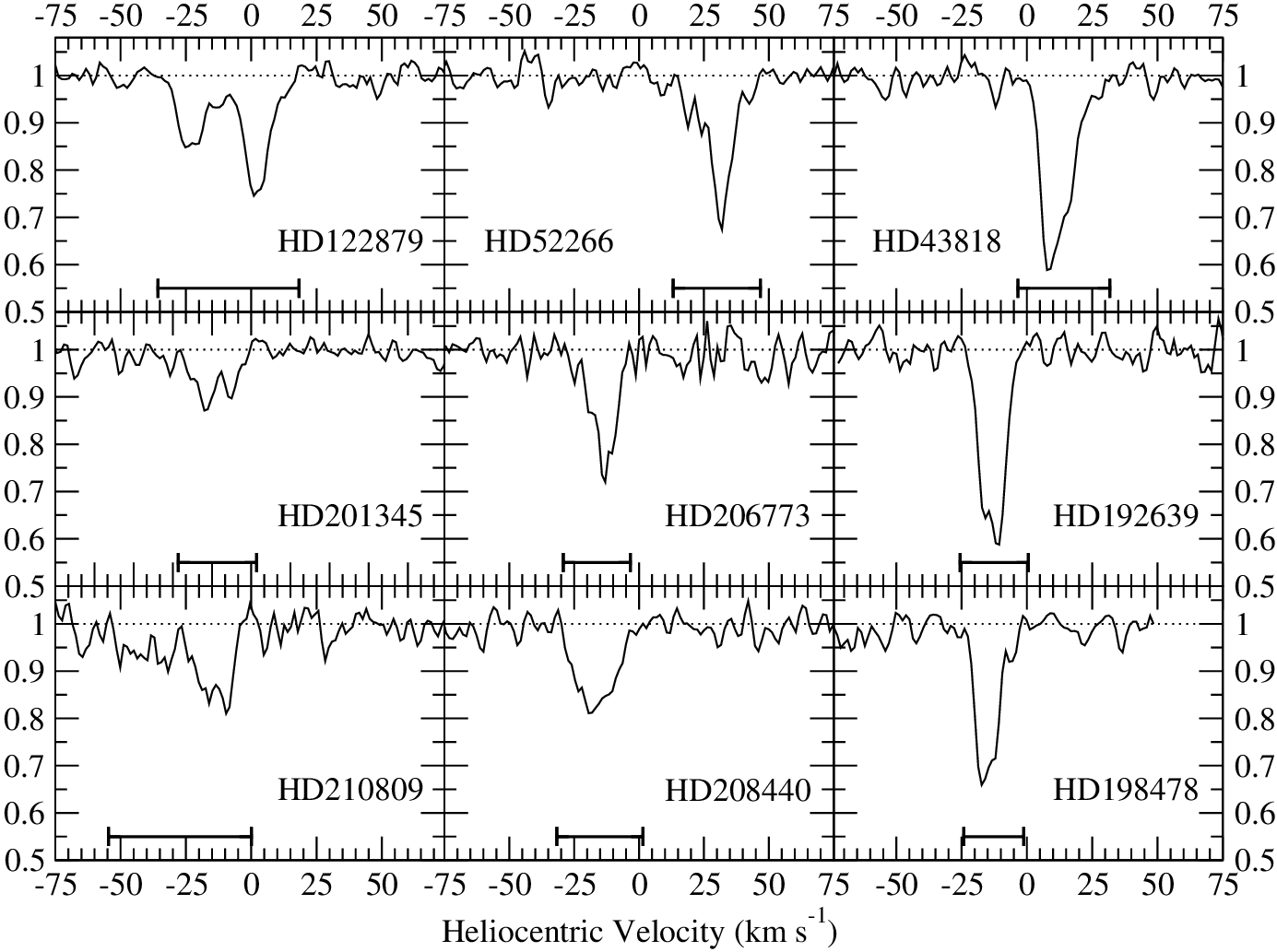}
\caption{\ion{O}{1} $\lambda$1356 Absorption Line Profiles\newline
A selection of \ion{O}{1} $\lambda$1356 absorption profiles are plotted above
as a function of the heliocentric velocity. In each panel, the velocity range
encompassed by the absorption features is identified by the line along the
bottom axis. These widths were determined in consideration of the profiles for
the dominant ionization states of oxygen, magnesium, phosphorus, manganese,
nickel, copper, and germanium; other lines used in an advisory role included
\ion{C}{1}, \ion{S}{1}, and \ion{Cl}{1} transitions in the 1170 -- 1372{\AA}
window. The upper left panel, corresponding to HD122879, depicts a typical
example of a multi-component oxygen profile. In the panels below it on the
left, for HD201345 and HD210809, examples of the least reliable measurements
are shown for relatively narrow and wide profiles, respectively. The middle
panels, for HD52266, HD206773, and HD208440, are profiles characteristic of the
general narrow (a few to one component) feature. Panels on the right are
representative of the deepest and most obvious oxygen profiles among the new
sight lines; the bottom right panel depicts the oxygen profile for HD198478,
which lies along the high-density plateau of the Boltzmann
function.\label{oprofs}}
\end{figure}

The new STIS observations of \ion{O}{1} $\lambda$1356 that are presented in this
paper were acquired in the course of two SNAPSHOT \textit{HST} programs. These
programs (GO8241, GO8662) were in part designed to study ISM cloud properties
as a function of distance and extinction by acquiring as many observations as
possible of O and B type stars that would produce useful spectra in 5, 10, or
20 minute exposures. The current sample includes results from all of these STIS
spectra for which a reliable oxygen abundance could be determined. All
exposures were made using the 0.2\arcsec$\times$0.2{\arcsec} STIS aperture with
either the E140H grating setup centered at 1271{\AA} or the E140M grating
centered at 1425{\AA}. The full STIS and \textit{FUSE} observation lists are
presented in Table~\ref{obstab}. Standard IRAF STSDAS reduction packages
were used in calibrating the STIS data, with the exception that X2DCORR rather
than SC2DCORR was applied to E140H exposures. This step was taken in order
to apply a data extraction code based on the procedure described by
\citet{how00}, used in the interest of consistency with the analysis of
\citet{car01}. In general, \ion{O}{1} $\lambda$1356 absorption appeared in the
overlap region between successive spectral orders, which were combined to
improve the S/N ratio; examples of the calibrated oxygen absorption profiles
are presented in Figure~\ref{oprofs}, including borderline cases as well as
features corresponding to much more reliable measurements. The final S/N per
pixel values near 1356{\AA} ranged from 30 -- 70.

Adhering to our established method \citep{car01,car03}, column densities were
determined from the \ion{O}{1} $\lambda$1356 absorption profiles using
independent apparent optical depth \citep{sav91} and profile fitting
\citep{mar95,wel91} procedures with the same constants applied as in previous
work (e.g., $f$ = 1.161$\times10^{-6}$). The results of both methods are
presented in Table~\ref{oxytab}. Of particular note, new column densities
derived using each method differ by less than 0.04 dex for an individual sight
line and these results are up to 0.13 dex larger than values derived under the
assumption of no saturation. The results of profile-fitting measurements for
new sight lines are adopted for all further analysis and are listed in
Table~\ref{oxyhydtab}, which also summarizes previous GHRS and STIS
measurements. Since the current analysis is comparative, the tabulated 1
$\sigma$ oxygen (and hydrogen) abundance errors do not include any contributions
from $f$-value uncertainties.

\begin{figure}
\plotone{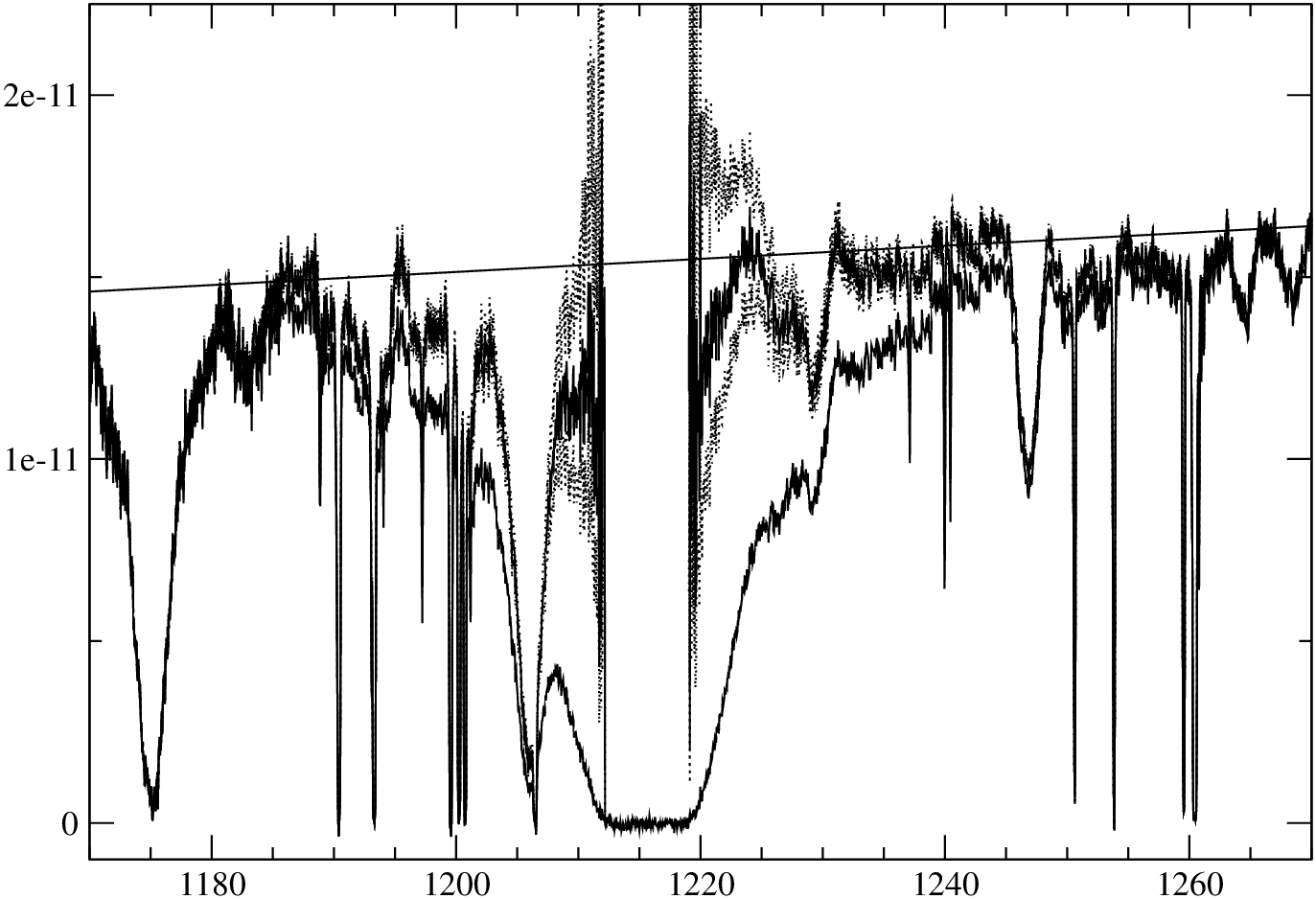}
\caption{A Typical Lyman-$\alpha$ Line Profile (HD165955)\newline
Measurement of the \ion{H}{1} column density is performed through division of
the calibrated and combined STIS spectrum near 1215.6700{\AA} (Ly$\alpha$) by
model absorption profiles generated by {FITS6P}. The STIS spectrum for HD165955
and the reconstructed continuum are shown above with solid lines; continuum
reconstructions for column densities differing from the best value by the
1 $\sigma$ uncertainty are plotted as dotted lines. These spectra have been
smoothed by a factor of 3 for presentation purposes. The straight line across
the spectrum is a linear approximation of the continuum in this
region.\label{h1fit}}
\end{figure}

For each sight line where the Lyman-$\alpha$ (Ly$\alpha$) absorption in the STIS
data appeared to be uncontaminated by a stellar contribution and \textit{FUSE}
data were available, the total interstellar hydrogen column density was
derived. The atomic measurement for each path was derived from STIS data in the
1160--1280{\AA} wavelength interval using a continuum-reconstruction method
\citep{boh75,dip94}. Generally, the current \ion{H}{1} measurements deviate
from values published in \citet{dip94} by less than 0.06 dex and even the few
severe discrepancies (up to 0.19 dex) are consistent within 1 $\sigma$
uncertainties; there are no systematic differences between our measurements and
those in the literature. A sample atomic hydrogen measurement is depicted in
Figure~\ref{h1fit}. Molecular hydrogen abundances were determined using $J$ = 0
and 1 transitions in the \textit{FUSE} spectral interval 1040--1120{\AA}, since
in general nearly 100\% of the molecular hydrogen along a sight line populates
these two rotational levels \citep{sav77,rac02}. This assumption was tested and
proved accurate for a few cases in the sample which exhibited prominent $J$ = 2
absorption features. These data were then calibrated using CALFUSE, the majority
with version 1.8.7. For nearly all of the sight lines, continuum-reconstruction
provided adequate abundance measurements; in cases where the column density was
too small to produce noticeable damping wings, an amount was determined using
the profile fitting code FITS6P\footnote{The FITS6P code models absorption
profiles based on varying the column densities, $b$-values, and velocities of
input interstellar components. More details on the technique can be found in
\citet{wel91}.}.

\begin{figure}
\plotone{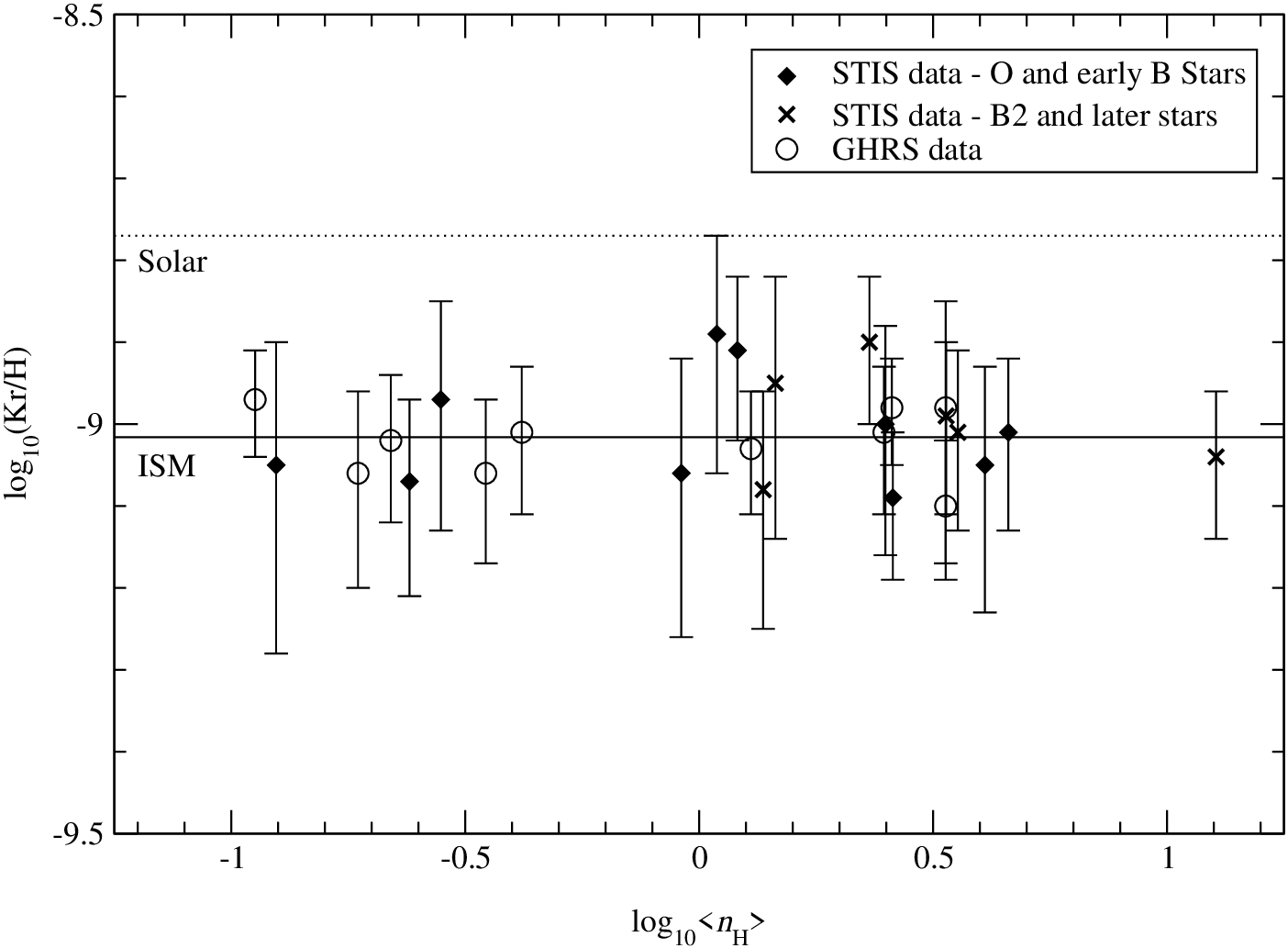}
\caption{Gas-phase Kr/H abundance ratios as a function of mean hydrogen sight
line density\newline
The variation of krypton abundance with mean total hydrogen sight line density
is depicted above for paths compiled from \citet{car03} and \citet{and03}.
The concern that the atomic hydrogen abundances derived for B3 stars in the
current sample suffer from stellar contamination is belied by the similarity
between the distributions of B3 Kr/H ratios and those for paths ending at stars
of other types for which such contamination is not a problem. HD116852 and
HD152590 have been excluded (see \citealt{car03}).\label{krplot}}
\end{figure}

While hydrogen column densities were readily determined for all sight lines
with both STIS and \textit{FUSE} data, contamination of the interstellar
Ly$\alpha$ profile by a stellar component was a concern for main sequence
stars with spectral types later than B1.5 or more luminous stars later than B2
\citep{dip94}. However, the interstellar krypton abundance can be used as a
secondary estimate of the total hydrogen column density for a sight line along
which it can be measured \citep{car97,car01}. In Figure~\ref{krplot}, krypton
and measured hydrogen data from the current sample \citep{car03} are combined
with measurements by \citet{and03} made for two stars that are distinct from
our collection. The uniformity of Kr/H ratios along sight lines toward stars
with types as late as B3 makes a strong argument for the position that the
\ion{H}{1} column densities determined toward these stars are reliable
tallies of their respective interstellar atomic hydrogen
abundances.\footnote{It should be noted that sight lines toward HD116852 and
HD152590 exhibit unusually large (near solar) Kr/H abundance ratios. These
paths are considered unique in this respect (see \citealt{car03}) and thus they
have been excluded from Figure~\ref{krplot}.} However, since a majority of them
are characterized by ${\langle}n_{\rm H}\rangle > 1.0~{\rm cm}^{-3}$ and they
comprise a significant portion of the denser sight lines being studied, we have
also assessed the degree of stellar contamination in $N$(\ion{H}{1})
measurements toward each of these later type stars.

Stellar contamination of the Ly$\alpha$ profile is potentially significant for
nine stars in the sample derived from direct measurements of both oxygen and
hydrogen abundances. In order to investigate the severity of this issue,
we have employed the \citet{dip94} procedure of using published values for each
star of the reddening corrected Str\"omgren [$c_1$] index ([$c_1$] = $c_1$ - [b
- y]) to estimate the stellar \ion{H}{1} component. For three stars, namely
HD79186 \citep{cra70}, HD165955 \citep{tob84}, and HD198478 \citep{cra73}, this
method implied that stellar contributions to their measured \ion{H}{1} column
densities were insignificant (less than 4\%). For two more stars, HD37367
\citep{wes82} and HD220057 \citep{sud92}\footnote{Direct Str\"omgren photometry
was unavailable for HD220057; Vilnius photometry and transforms published by
\citet{kal02} were used to derive $c_1$.}, the suggested stellar fractions rise
to about 15\%; given this level of contamination, the molecular hydrogen
abundances determined for these sight lines limit any possible shift in total
abundance to about 0.06 dex, less than the uncertainty associated with the
current hydrogen measurements. Published photometry for HD203532 \citep{egg77}
and HD212791 \citep{cam66} imply upper limits on the stellar component of 50
and 25\%, respectively, of the total measured atomic hydrogen abundances for
these paths. In light of some uncertainty in the index for HD203532 and the
excellent agreement between Kr/H abundance ratios for these paths derived
without a stellar correction applied and the interstellar mean, however, the
atmospheric contribution to $N$(\ion{H}{1}) for these paths may not be as
severe is the photometry indicates. We must also emphasize that the adjustment
of these two O/H datapoints by even their maximal shifts incurs only a minor
change in our results (less than 0.02 dex in the plateaus defined in
Section~\ref{section_odepletion_effects}).

Among the stars potentially affected by stellar \ion{H}{1} absorption, HD27778
and HD147888 are crucial because the inferred ${\langle}n_{\rm H}\rangle$
values are high. Toward HD27778, \citet{rac02} estimated the interstellar
\ion{H}{1} column density using $E(\bv)$; their results for both $N$(\ion{H}{1})
and $N$(H$_2$) agree with ours within error and the derived total hydrogen
column densities differ by only 0.02 dex. Furthermore, since its Kr/H abundance
ratio almost precisely matches the interstellar mean, we conclude that the
HD27778 hydrogen measurement is not significantly influenced by a stellar
\ion{H}{1} component. The hydrogen abundance toward HD147888 ($\rho$ Oph D) was
taken from \citet{car01}, who used HD147933 ($\rho$ Oph A) as a proxy.
Specifically $N($H$_{total})$ toward HD147933 was derived from the weighted
mean of atomic measurements published by \citet{boh78} and \citet{dip94} and
the molecular abundance reported by \citet{sav77}. It is important to note that
\citet{dip94} found that the HD147933 Ly$\alpha$ profile was essentially
uncontaminated by stellar absorption and that the atomic hydrogen abundance
determined from the HD147888 spectrum matches the HD147933 weighted mean
precisely.

In summary, a detailed investigation of atomic hydrogen diagnostics for these
stars suggests that the current values are reasonable \ion{H}{1} measurements
and strengthens the case for using krypton as a reliable indicator of the
total interstellar hydrogen abundance for sight lines lacking accurate and
direct determinations. Accordingly, \ion{Kr}{1} $\lambda$1236 absorption
features have been used to derive the total hydrogen abundance toward HD43818,
HD148594, and HD175360; \ion{H}{1} toward the latter two stars is severly
overestimated by the Ly$\alpha$ profile and the lack of a \textit{FUSE}
observation of HD43818 leaves $N$(H$_2$) undetermined for that sight line.
Sight line properties not related to hydrogen have been included in the
Appendix (Table~\ref{apptab}) for all of the compiled STIS and GHRS paths.

\section{Density-Dependent Depletion}
\label{section_odepletion_effects}
\begin{figure}
\plotone{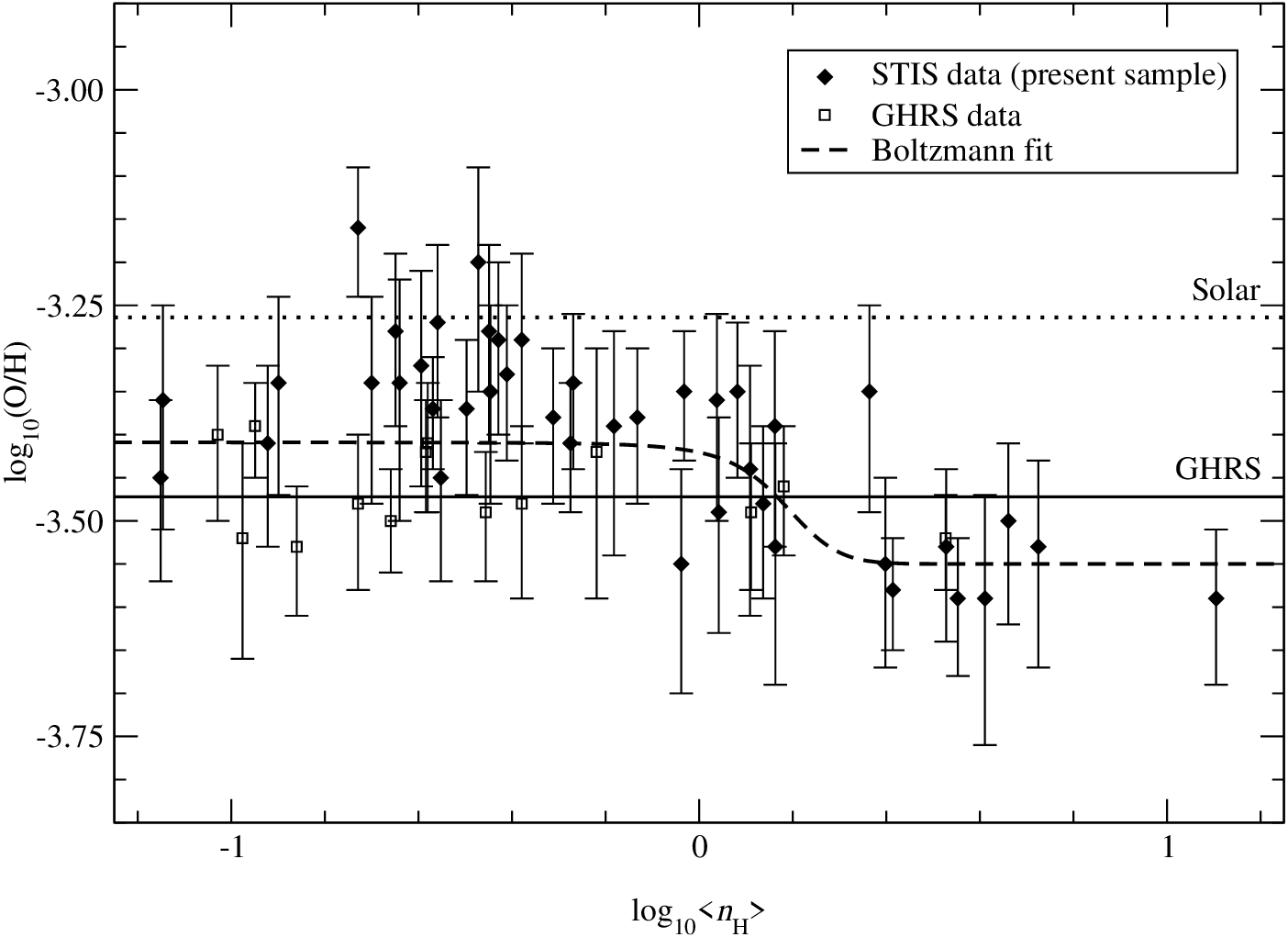}
\caption{Gas-phase O/H abundance ratios as a function of mean hydrogen sight
line density\newline
Oxygen abundances are plotted above as a function of the sight line property
${\langle}n_{\rm H}\rangle$, identifying a trend consistent with the
density-dependent depletion signatures evident for other elements \citep{car04}.
Fit with a 4-parameter Boltzmann function (dashed line), the data indicate that
gas-phase oxygen levels vary from 390$\pm$10 to 284$\pm$12 $ppm$ as
${\langle}n_{\rm H}\rangle$ increases from 0.1 to 10.0 cm$^{-3}$. The solar
\citep{hol01} and GHRS \citep{mey98} O/H levels are plotted for reference; the
\citet{asp04} solar oxygen abundance is somewhat lower than the \citet{hol01}
value, at log$_{10}$(O/H)$_\sun$ = $-$3.34$\pm$0.05 (457$\pm$56 $ppm$), but the
two levels agree within error.\label{oboltz}}
\end{figure}

Previous studies of the variation of interstellar elemental abundances have
investigated the dependence they exhibit on several sight line properties
including a variety of extinction measures and the fraction of molecular
hydrogen \citep{jen86,van88}. The characteristic producing the cleanest signal
in measured gas-phase abundances is the sight line mean of total hydrogen
column density, ${\langle}n_{\rm H}\rangle$; specifically, decreasing gas-phase
abundances (or increasing inferred depletions onto dust) are positively
correlated with increasing ${\langle}n_{\rm H}\rangle$. Interstellar oxygen,
however, did not exhibit the same dependence as other elements---in fact, the
majority of UV absorption studies to date have found a remarkable O/H ratio
consistency within their samples (e.g., \citealt{yor83,kee85,mey98,and03}). But
given the large amounts of CO and other oxygen-bearing molecules observed in
dense clouds and conditions that are favorable for the growth of dust grains,
it has been expected that oxygen depletion must be enhanced as the particle
density increases to values achieved in molecular clouds. The first observations
that suggested such enhancement were presented by \citet{car01}; these data are
included in the current sample and provide the basis for a functional
description of the interstellar gas-phase O/H abundance ratio based on
${\langle}n_{\rm H}\rangle$ (see Figure~\ref{oboltz}).

\begin{figure}
%\plotfiddle{plotdir/fig5.eps}{0in}{0}{425}{425}{16}{0}
\plotone{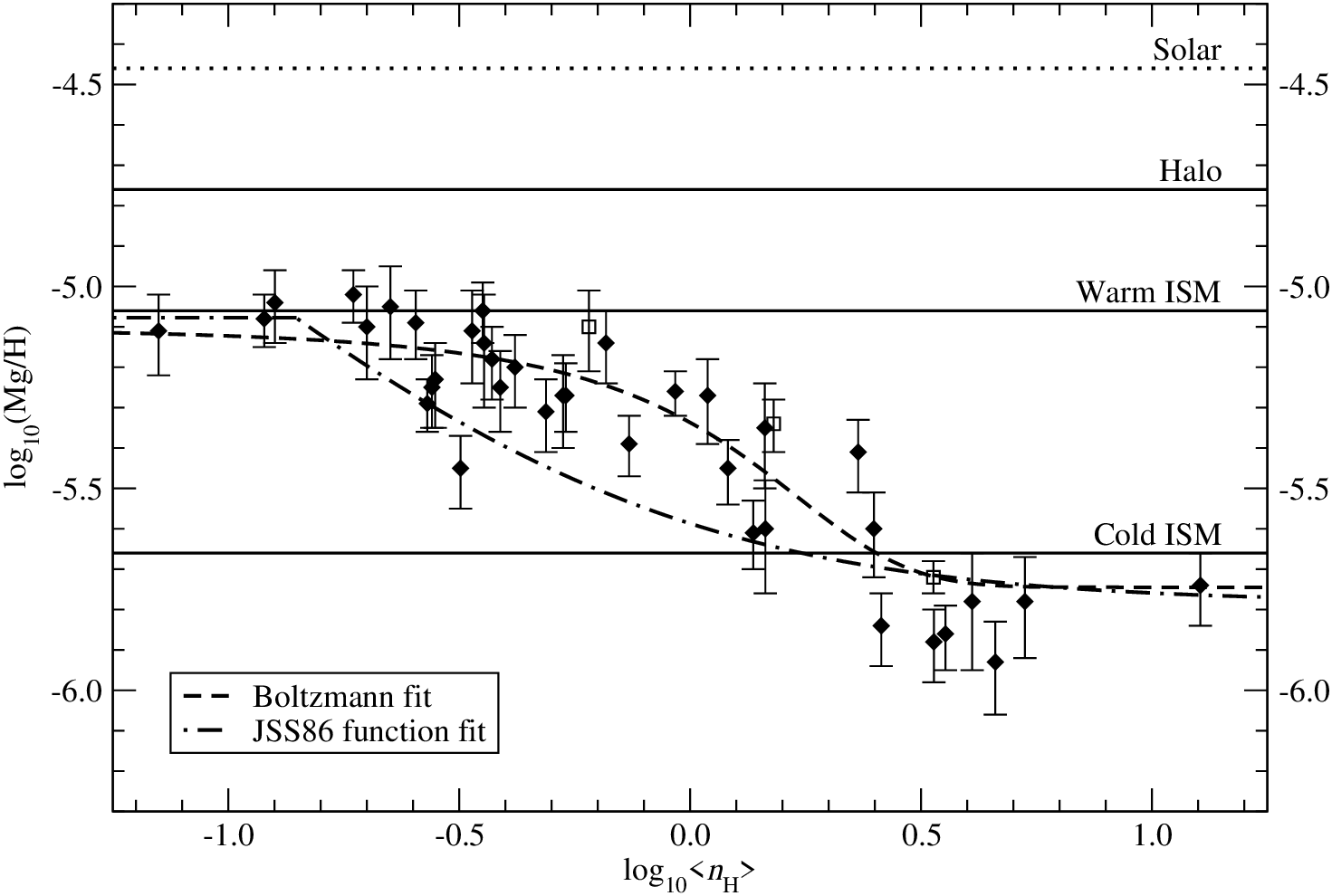}
\caption{Gas-phase Mg/H abundance ratios as a function of mean hydrogen sight
line density\newline
Gas-phase magnesium to hydrogen abundance ratios from \citet{car04} are plotted
above as a function of the mean total hydrogen sight line density. Visual
inspection of this graph and the ${\langle}n_{\rm H}\rangle$ dependency
depicted in Figure~\ref{oboltz} demonstrates how similar they are in form, as
the gas-phase abundance undergoes a smooth transition from one apparent plateau
to a second lower one as mean sight line density increases. Furthermore, the
4-parameter Boltzmann function appears better suited to the data than the
2-parameter function used by \citet{jen86} (identified as JSS86 in the plot
above). The solar reference ratio was taken from \citet{hol01} and the halo and
warm and cold ISM Mg/H levels were compiled by \citet{wel99}.\label{mgboltz}}
\end{figure}

The 4-parameter Boltzmann function overlaid on the data in Figure~\ref{oboltz}
has the functional form: \begin{equation} ({\rm O/H})_{\rm gas} = ({\rm O/H})_c
+ \frac{({\rm O/H})_w - ({\rm O/H})_c}{1 + e^{({\langle}n_{\rm H}\rangle -
{\langle}n_0\rangle)/m}}, \end{equation} where $({\rm O/H})_w$ and $({\rm
O/H})_c$ are the oxygen gas-phase abundance levels appropriate for low and high
mean sight line densities, respectively, ${\langle}n_0\rangle$ is the
inflection point of the curve, and $m$ is its slope at the inflection point.
This function is a variation of the form used by \citet{jen86}, where $\left(1
+ e^{({\langle}n_{\rm H}\rangle - {\langle}n_0\rangle)/m}\right)^{-1}$ replaces
the original factor $n_{\rm sw}/{\langle}n_{\rm H}\rangle$. The earlier form
has only two free parameters (the two depletion levels), under the assumption
that $n_{\rm sw}$ (0.10 cm$^{-3}$; \citealt{jen86}) is a canonical density for
the warm neutral ISM below which no cool diffuse clouds contribute to the bulk
of interstellar material along a sight line. As shown in Figure~\ref{mgboltz},
the revised form eliminates the kink at $n_{\rm sw}$ and appears to follow the
data for oxygen and other elements more naturally (Figure~\ref{mgboltz};
\citealt{car04}); hence, it is used in this analysis.

Although ${\langle}n_{\rm H}\rangle$ is a somewhat crude measure of the
environment through which a sight line passes [${\langle}n_{\rm H}\rangle$
$\equiv$ $N$(H)/$d_\ast$], its elemental abundance response
can be interpreted in terms of the idealized ISM framework postulated by
\citet{spi85}. In this model, the superposition of depletion signatures from
three phases of the ISM convolved with the distribution of each phase's filling
factor as a function of the mean sight line density\footnote{According to the
\citet{spi85} model, warm neutral material dominates sight lines with
${\langle}n_{\rm H}\rangle < 0.2$ cm$^{-3}$, cool diffuse clouds are prevalent
where ${\langle}n_{\rm H}\rangle \approx 0.7$ cm$^{-3}$, and denser clouds
contribute the majority of matter along paths with ${\langle}n_{\rm H}\rangle >
3.0$ cm$^{-3}$.} leads to a clear dependence of gas-phase abundance on
${\langle}n_{\rm H}\rangle$. An implicit assumption of the model is that each
ISM phase exhibits a characteristic depletion level. Although densities vary in
a continuous fashion in the natural ISM, the appearance of a plateau in a plot
of abundance vs. ${\langle}n_{\rm H}\rangle$ would indicate that an element's
depletion is fairly constant over a range of real densities. Consequently,
Figure~\ref{oboltz} suggests that oxygen depletion remains roughly fixed in two
distinct interstellar density intervals.

\begin{figure}
\plotfiddle{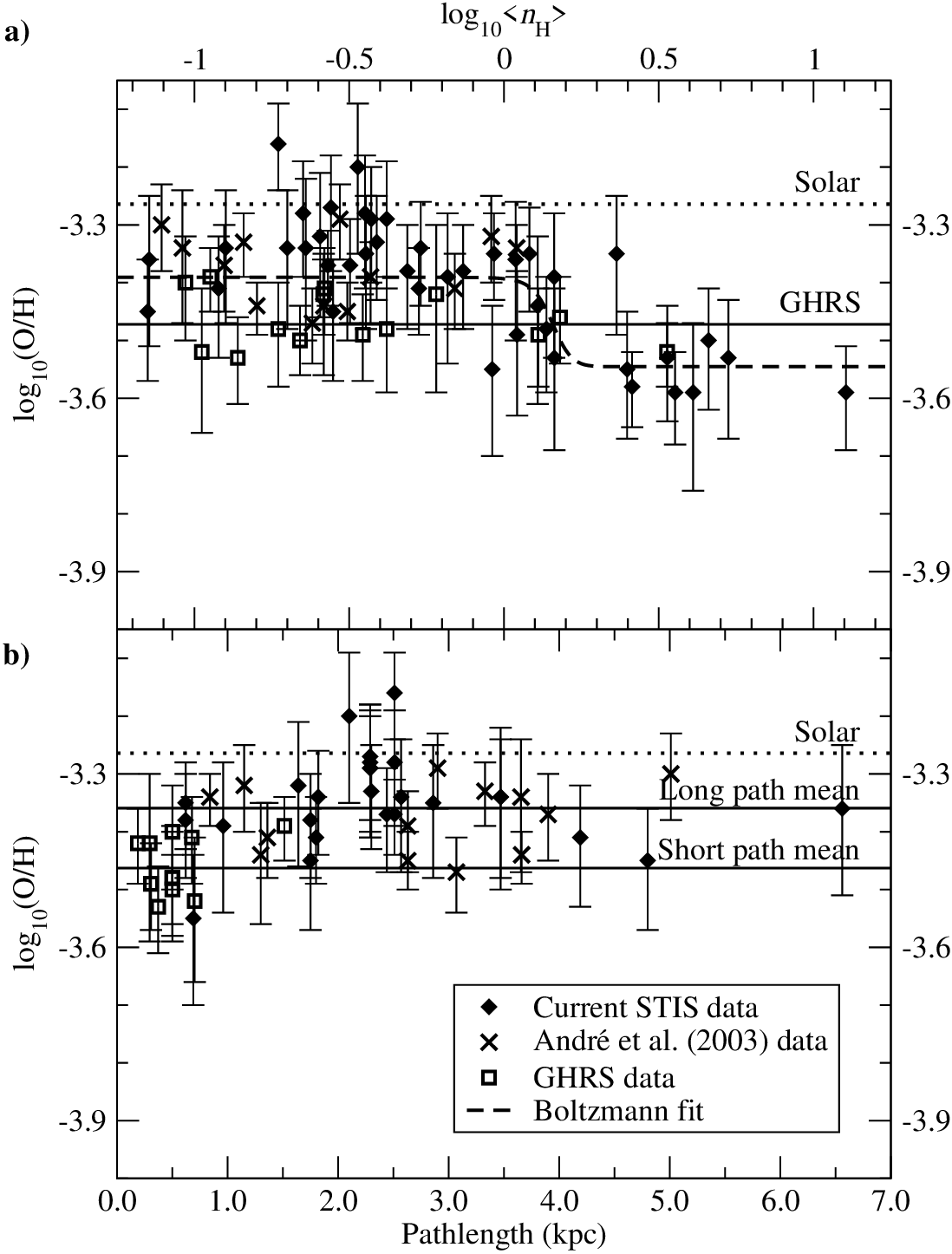}{0in}{0}{425}{425}{16}{0}
%\plotone{plotdir/fig6.eps}
\caption{STIS and GHRS Gas-phase O/H abundance ratios as functions of sight
line pathlength and mean hydrogen sight line density\newline
{\bf a)} Data from \citet{and03} are added to the comparison of STIS and GHRS
O/H abundance ratios as a function of ${\langle}n_{\rm H}\rangle$. The
additional data are consistent with our measurements for low density sight
lines and do not significantly alter the parameters determined for a Boltzmann
function fit to the entire dataset. {\bf b)} Gas-phase O/H ratios for paths
with ${\langle}n_{\rm H}\rangle$ $<$ 1.0 cm$^{-3}$ are plotted as a function of
distance to the target star, with the addition of \citet{and03} data. The GHRS
datapoints generally have lower O/H ratios than their STIS counterparts,
producing a 0.10 dex difference between the mean O/H value for sight lines
shorter and longer than 800 pc. The solar \citep{hol01} and GHRS \citep{mey98}
O/H levels are plotted for reference in each panel; the \citet{asp04} solar
oxygen abundance is somewhat lower than the \citet{hol01} value, at
log$_{10}$(O/H)$_\sun$ = $-$3.34$\pm$0.05 (457$\pm$56 $ppm$).\label{hdist}}
\end{figure}

%In addition, since this function describes a mean depletion signature, the relative data scatter provides a rough measurement of the inherent variability of the total (gas + dust) O/H abundance ratio.
By performing an uncertainty-weighted fit of the Boltzmann function to the data,
reliable warm- and cold-ISM depletion levels can be measured simultaneously.
Using the full oxygen data sample, the values derived from the entire sample
correspond to levels of log$_{10}$(O/H) = $-$3.41$\pm$0.01 (390$\pm$10 $ppm$)
for warm, low-${\langle}n_{\rm H}\rangle$ sight lines and log$_{10}$(O/H) =
$-$3.55$\pm$0.02 (284$\pm$12 $ppm$) along paths dominated by cooler and denser
material with the transition centered at ${\langle}n_{\rm H}\rangle$ =
1.50$\pm$0.22 cm$^{-3}$. The fit to the data also indicates that the number of
parameters associated with the chosen Boltzmann function is reasonable;
$\chi_\nu^2$ = 0.69 and the relative data scatter is 0.09 dex. Although the
present discussion is focussed on reconciling our STIS data with GHRS results,
supplementing the current data set with those sight lines unique to
\citet{and03} does not alter these determinations significantly
(log$_{10}$(O/H)$_{warm}$ shifts to $-$3.40, log$_{10}$(O/H)$_{cold}$ is
unchanged, the inflection point occurs at 1.46 cm$^{-3}$, and $\chi_\nu^2$
rises to 0.75; see Figure~\ref{hdist}a). Scatter in the data includes
contributions from both intrinsic variability and measurement uncertainty,
however the apparent distribution of the data around the Boltzmann function has
a width roughly equal to a typical datapoint error bar (0.09 dex). But since
the Boltzmann function represents a mean depletion signature, a portion of the
data dispersion is due to the spread of gas/dust fractions allowed at a given
value of ${\langle}n_{\rm H}\rangle$. In the idealized picture of
\citet{spi85}, this would correspond to scatter in the proportion of sight line
material contributed by each of the ISM phases. In consideration of the
approximate parity between the data dispersion around the depletion signature
and estimated O/H ratio uncertainties, the uniformity of ISM elemental
abundances implied for length scales of several hundred parsecs by gas-phase
Kr/H ratios \citep{car03} might be extended to distances of a few kpc using
oxygen.

The plateaus determined from the Boltzmann fit not only define the mean
gas-phase oxygen abundances for sight lines of a given ${\langle}n_{\rm
H}\rangle$, they also bear on our understanding of dust composition. The
sight lines examined in this paper will not contain significant amounts of ice.
Any depletion of an element, therefore, should be into grains. The most likely
oxygen carriers in dust are silicates and oxides; thus, the large cosmic
abundances and heavy depletions of magnesium and iron suggest that the majority
of the O-bearing grains will be associated with these two elements. Thus,
silicon, magnesium, and iron depletions can be used to place limits on the
amount of oxygen incorporated into grains. Our discussion of dust composition
assumes that solar photospheric abundances well represent the total (gas +
dust) abundances in the ISM; specifically, we use the solar abundances of
silicon, magnesium, and iron from \citet{hol01}.

The fractions of silicon and magnesium in dust trace each other extremely well
from diffuse halo environments to the denser warm and cool disk regions
\citep{sem96,sav96,the99,sof04a}. Moreover, since their solar abundances are
nearly identical, approximately equal numbers of silicon and magnesium atoms
are incorporated into dust from low- ($\lesssim$50\% of silicon resides in
dust) to high-depletion interstellar regions ($\gtrsim$90\% of silicon is in
dust; \citealt{fit97}). In combination with the fact that Mg-bearing silicates
better fit interstellar spectral features than Fe-bearing silicates
\citep{whi03}, this correlation suggests that silicon and magnesium are usually
incorporated into the same grains. \citet{jon00}, however,
finds that silicon is returned to the gas-phase ISM a bit more readily
than magnesium, so that not every atom of one element in the dust will be
paired with one atom of the other. Nevertheless, if we assume that most
magnesium atoms deplete onto silicate grains in a 1-to-1 ratio with silicon
atoms, then a likely grain mineral would be MgSiO$_{3}$ \citep{oss92} which has
a high O-to-Si ratio. If the silicates are primarily Mg-based, then most or all
of the iron would have to be incorporated into metal grains, or more likely
oxides such as FeO, Fe$_{3}$O$_{4}$ or Fe$_{2}$O$_{3}$ \citep{jon90}. Using
these grain types, an average halo-like region will have up to 72$\pm$14
oxygen atoms incorporated into dust per million hydrogen atoms in the gas,
while for typical cool disk regions the value would rise to 140$\pm$14 in the
same units. It should be noted that the uncertainties have been derived from
the solar abundances alone and do not account for fluctuations of the measured
gas-phase abundances in these regions. Other reasonable combinations of
silicate and oxide grain types would all suggest lower oxygen dust abundances.

%Asplund et al. (2004) says 457+/-56 ppm => 67$\pm$58 (v.good agreement) for
%low n_H and 175$\pm$59 (good agreement).

Two recent measurements of the solar photospheric oxygen abundance are
(O/H)$_\sun$ = 545$\pm$107 and 457$\pm$56 $ppm$ (\citealt{hol01} and
\citealt{asp04}, respectively); a weighted average yields (O/H)$_\sun$ =
476$\pm$50 $ppm$. Assuming this abundance for the ISM, our low- and high-density
sight line averages imply 86$\pm$51 and 192$\pm$51 oxygen atoms incorporated
into dust per million hydrogen atoms in the gas phase, respectively. The
dust-phase oxygen abundance in the average low-density sight line would thus
agree quite well with our above halo estimate within the given uncertainties,
although the corresponding high density path value is just at the edge of 1
$\sigma$ larger than the value we have derived for cool disk material. Opting
for the \citet{asp04} abundance rather than the mean would somewhat improve the
accord in both regimes; however, given the magnitude of the uncertainties
associated with the \citet{hol01} and \citet{asp04} solar references and recent
work suggesting that these levels are inconsistent with helioseismic results
and current opacity tables \citep{bas04}, we regard none of these values as
strictly preferred. Upon adopting any of these photospheric
standards, the gas-phase abundance levels we measure imply 106$\pm$16 more
oxygen atoms per million hydrogen atoms in the gas along high mean density
sight lines than along more diffuse paths. This difference is somewhat larger
than would be expected solely from an enhanced oxygen dust abundance
(68$\pm$20). The gap is thus difficult to explain physically unless oxygen in
denser regions can be incorporated into grain types other than those discussed
or if it is substantially depleted into molecules. Of note, CO is not a major
contributor to the carbon abundance along 5 of the denser sight lines in this
sample \citep{sof04b}.

\section{Local vs. Distant O/H Ratios}
\label{section_heliodist_effect}
The large inferred oxygen dust abundance for high-${\langle}n_{\rm H}\rangle$
sight lines might be influenced, however, by an apparent spatial effect. GHRS
sight lines were the basis for the conclusion that the oxygen abundance in the
local ISM was constant, in spite of the diversity in the properties of paths
included in the study \citep{mey98}. Yet gas-phase O/H ratios determined from
STIS and \textit{FUSE} data imply measurably higher abundances among paths with
low mean densities (${\langle}n_{\rm H}\rangle$ $<$ 1.0 cm$^{-3}$) than were
measured using GHRS\footnote{Hydrogen column densities toward stars observed
with GHRS were derived a combination of \textit{Copernicus} \citep{sav77,boh78}
and \textit{International Ultraviolet Explorer} \citep{dip94} measurements.
Atomic hydrogen column densities determined from each instrument's data were
consistent within 1 $\sigma$ errors and their weighted mean was used for each
GHRS sight line in this sample.}. The property that aside from the O/H ratio
most clearly distinguishes the samples observed by GHRS, STIS, and
\textit{FUSE} is sight line pathlength. As shown in Figure~\ref{hdist}b, the
GHRS sample was constructed almost exclusively from sight lines shorter than
about 800 pc while STIS has generally examined more distant targets.
Consequently, if the GHRS and STIS low density (${\langle}n_{\rm H}\rangle$ $<$
1.0 cm$^{-3}$) samples are divided into groups shorter and longer than 800 pc,
the weighted mean O/H abundance ratios diverge (log$_{10}$(O/H)$_{\rm short}$ =
-3.46$\pm$0.02 and log$_{10}$(O/H)$_{\rm long}$ = -3.36$\pm$0.02).

\begin{figure}
\plotone{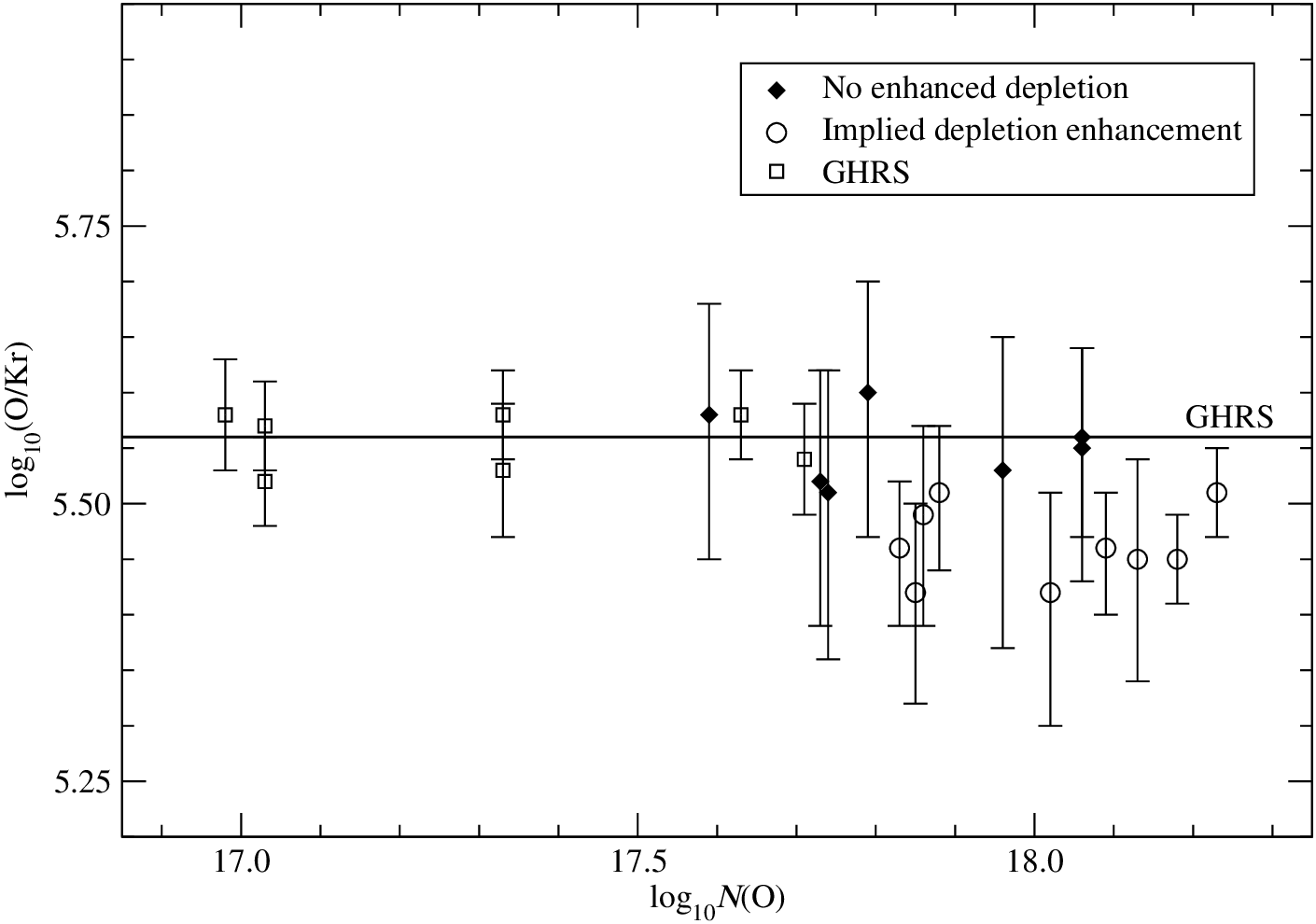}
\caption{Gas-phase O/Kr abundance ratios as a function of gas-phase oxygen
abundance\newline
Gas-phase O/Kr abundance ratios are plotted above for all paths in the current
sample with reliable oxygen and krypton measurements except HD116852 and
HD152590; the krypton abundances along these two sight lines are unusually
large and are not included for this reason. The GHRS data are all consistent
with a single mean ratio and are well-matched by lower density STIS sight
lines. However, higher density paths that exhibit enhanced oxygen depletion
with respect to hydrogen also evince lower O/Kr ratios, in contrast to the GHRS
data. Unfortunately, this distinction is not definitive because of the large
uncertainties associated with the STIS measurements.\label{okr}}
\end{figure}

Perhaps the simplest explanation for any short and long sight line abundance
discrepancy would be that the local ISM has experienced an infall of metal-poor
gas, recently enough that mixing processes have not yet erased its elemental
abundance signature \citep{mey94}. However, since all but one of the
high-${\langle}n_{\rm H}\rangle$ paths are also shorter than 800 pc, they would
also be affected by infall. In this scenario, the density-depletion signature
that is evident in the current data would be mitigated but not erased, since it
is apparent in Figure~\ref{okr} that sight lines that exhibit enhanced oxygen
depletion when compared with hydrogen also evince somewhat reduced gas-phase
O/Kr ratios. As a result, metal-poor gas infall might conveniently account for
the larger-than-anticipated oxygen depletion enhancement that we determine for
high mean density sight lines. Such infall, however, should also reduce other
gas-phase elemental abundance ratios by a similar factor---but although spatial
variations in Cu/H ratios for a similar grouping of sight lines are suggestive
\citep{car04}, not enough measurements of sufficient quality have been made to
rigorously evaluate this scenario. The uniformity of Kr/H for many of the same
sight lines that appear in the current sample \citep{car03} does not provide
clear evidence for or against this possibility, since the overwhelming majority 
of these paths are shorter than 800 pc.

\begin{figure}
\plotone{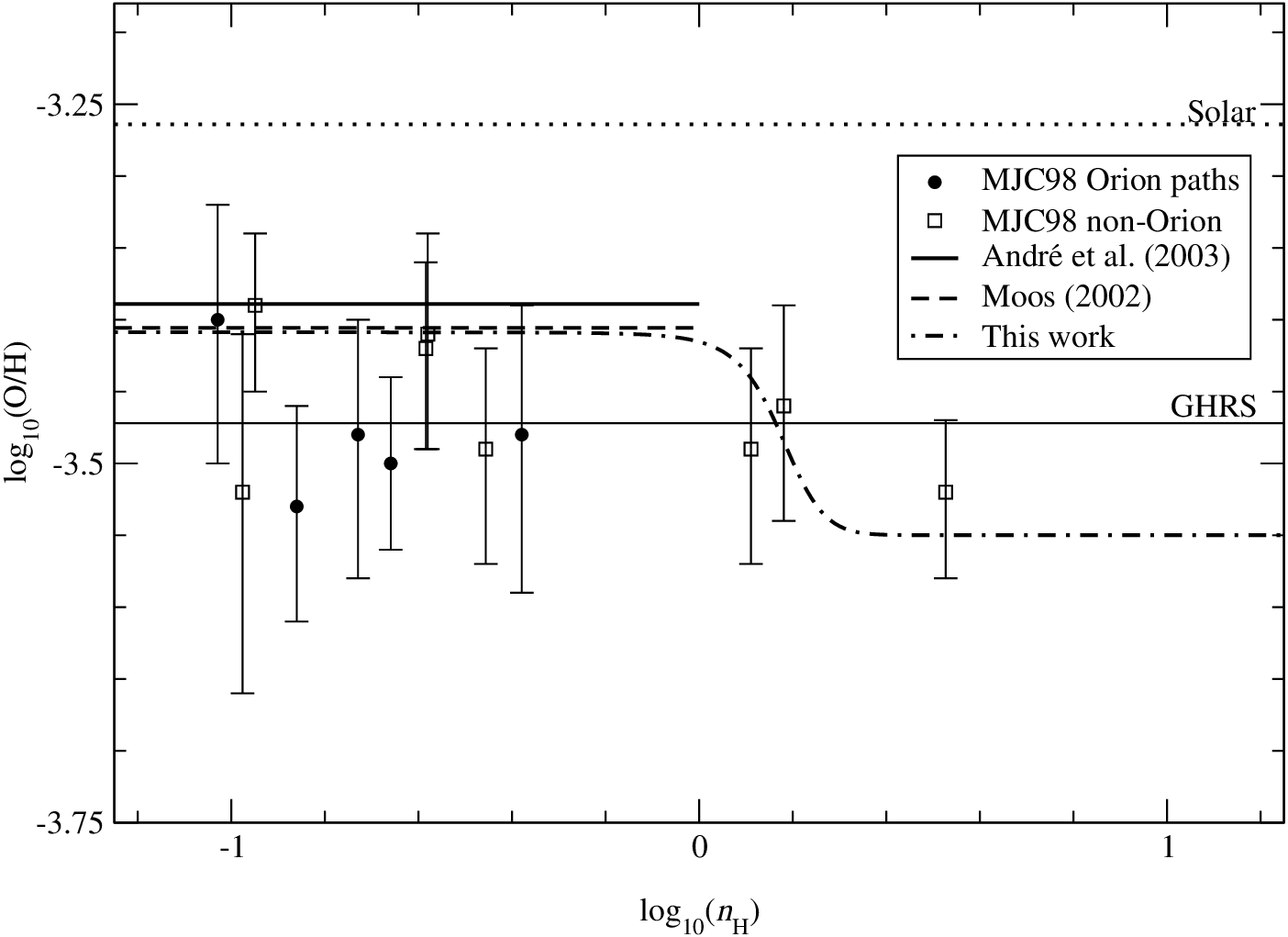}
\caption{GHRS gas-phase O/H abundance ratios\newline
Gas-phase O/H abundance ratios for the GHRS sight lines compiled by
\citet{mey98} (identified in the plot as MJC98) are plotted above as a function
of mean total hydrogen sight line density. Reference oxygen levels are the GHRS
weighted mean \citep{mey98}, the solar ratio \citep{hol01}, the \citet{and03}
and \citet{moo02} low-${\langle}n_{\rm H}\rangle$ means and the
density-dependent depletion trend identified in this paper. Of the five Orion
sight lines appearing in \citet{mey98}, only $\iota$ Ori is situated near the
bulk of other low-${\langle}n_{\rm H}\rangle$ oxygen measurements. The others
dominate the sample that results in the reduced mean for short diffuse sight
lines that distinguish them from longer diffuse paths.\label{shortlown}}
\end{figure}

An intriguing alternative gap explanation arises, however, if one examines the
character of the short diffuse sight lines exhibiting low O/H ratios (see
Figure~\ref{shortlown}). These GHRS sight lines are dominated by paths directed
toward Orion; among the five \citet{mey98} sight lines pointed in this
direction, four fall at or below the GHRS weighted O/H mean and only one sits
near the gas-phase oxygen levels determined for diffuse paths using STIS and
\textit{FUSE} data. It has previously been suggested that the Orion region is
oxygen-poor; for instance, \citet{cun94} found that stellar atmospheres there
were consistent with Orion nebular measurements, approximately 0.26 dex
underabundant in oxygen relative to the Sun. While subsequent downward
revisions of the solar level have reduced the contrast, it should be noted that
these independent oxygen abundance values agree with the Orion GHRS
measurements. Although one would expect that mixing processes would alleviate
the signature of oxygen-poor stars in Orion interstellar gas studied at the
current epoch, the coincidence of these measurements suggest that the apparent
gap between short and long low-density sight line oxygen levels might be a
consequence of reduced abundances in Orion convolved with measurement
uncertainties.

Regardless of the origin of the pathlength-based gap among low-${\langle}n_{\rm
H}\rangle$ sight lines, the effect on the depletion signature determined by
fitting the oxygen data to the Boltzmann function is limited in magnitude. A
sample constructed of dense sight lines and only the longer diffuse paths shifts
the low-density plateau by just 0.03 dex, raising the gas-phase O/H ratio by
about 8\%; these values are not significantly altered if the \citealt{and03}
data for sight lines distinct from our sample are also included. In light of
the very minor adjustments that this effect generates in the depletion
signature, it should be noted that although the trends identified in the
current set of oxygen data are clearly recognizable, they are present only at
levels comparable to datapoint uncertainties. Consequently, additional precise
interstellar gas-phase abundance measurements (oxygen \textit{and} hydrogen),
for both low and high mean density sight lines, are required in future in order
to more reliably constrain our knowledge of the interstellar oxygen abundance
in the Milky Way, the reasons why the O/H ratio may vary spatially, and the
role oxygen plays in dust.

\acknowledgments
We would like to thank the referee for suggesting amendments that have improved
the rigor of our results, and acknowledge that this research has made use of
the SIMBAD database, operated at CDS (Strasborg France), and the on-line
General Catalogue of Photmetric Data (University of Lausanne, Switzerland;
\citealt{mer97}).

\appendix
\section{APPENDIX}
\label{section_appendix}
The common sight line properties for stars in the current sample are presented
in Table~\ref{apptab} for easy reference. Spectral types were compiled by James
T. Lauroesch in preparation for \textit{HST} SNAPSHOT programs GO8241 and
GO8662; the tabulated color excesses and distances (assigned by group
membership) are drawn from \citet{ber58}, \citet{boh81}, \citet{hum84},
\citet{sea84}, \citet{wes85}, \citet{mer86}, \citet{gie87}, \citet{ree93},
\citet{bro94}, \citet{dip94}, \citet{sem96}, \citet{rab97}, \citet{egg98},
\citet{dez99}, \citet{bau00}, \citet{ree00}, and \citet{rac02}.
%Include the fourth table "Summary of Sight Line Properties" \ref{apptab} here.

\clearpage

\clearpage

\clearpage

\clearpage

\clearpage

\clearpage

%\begin{figure}
%\plotfiddle{plotdir/fig6.eps}{0in}{0}{425}{425}{16}{0}
%\plotone{plotdir/fig6.eps}
%\caption{Andr\'e et al. added\newline
%As in Figure~\ref{oboltz}, oxygen abundances have been plotted above as a
%function of the sight line property ${\langle}n_{\rm H}\rangle$, identifying a trend consistent with the density-dependent depletion signatures evident for other elements \citep{car04}.  Fit with a 4-parameter Boltzmann function (dashed line), the data indicate that gas-phase oxygen levels vary from 390$\pm$15 to 284$\pm$18 $ppm$ as ${\langle}n_{\rm H}\rangle$ increases from 0.1 to 10.0 cm$^{-3}$. The solar \citep{hol01} and GHRS \citep{mey98} O/H levels are plotted for reference; the \citet{asp04} solar oxygen abundance is slightly lower than the \citet{hol01} value, at log$_{10}$(O/H)$_\sun$ = $-$3.31$\pm$0.05 (490$\pm$60 $ppm$).\label{oboltz+and}}
%\end{figure}

%\clearpage

\clearpage

\clearpage

\clearpage

\begin{deluxetable}{lcccccc}
%\rotate
\tablecaption{Summary of Observations \label{obstab}}
\tablewidth{0pt}
\tablehead{
&\colhead{STIS}&&\colhead{Time}&\colhead{\textit{FUSE}}&&\colhead{Time}\\
Sight Line&\colhead{Data Set}&\colhead{Date}&\colhead{(s)}&\colhead{Data Set}&
\colhead{Date}&\colhead{(s)}\\
}
\startdata
HD1383   & O5C07C010 & 1999 Nov 08 & 1440 & B0710101000 & 2000 Sep 30 &23332 \\
HD12323  & O63505010 & 2001 Apr 02 & 1440 & P1020202000 & 1999 Nov 25 & 3865 \\
HD13268  & O63506010 & 2001 Feb 01 & 1440 & P1020304000 & 1999 Nov 24 & 4438 \\
HD14434  & O63508010 & 2001 Apr 06 & 1440 & P1020504000 & 1999 Nov 24 & 4441 \\
HD36841  & O63516010 & 2000 Nov 24 & 1440 &     ---     &     ---     & ---  \\
HD37367  & O5C013010 & 2000 Mar 31 & 360  &     ---     &     ---     & ---  \\
HD43818  & O5C07I010 & 2001 Apr 10 & 1440 &     ---     &     ---     & ---  \\
HD52266  & O5C027010 & 2000 Mar 11 & 360  &     ---     &     ---     & ---  \\
HD63005  & O63531010 & 2001 May 07 & 1440 & P1022101000 & 2000 Apr 05 & 5311 \\
HD71634  & O5C090010 & 2000 May 22 & 720  & Z9012201000 & 2002 Mar 08 & 504  \\
HD72754  & O5C03E010 & 1999 Oct 14 & 720  & B0710302000 & 2002 May 04 & 5950 \\
HD79186  & O5C092010 & 2000 May 25 & 360  & P2310301000 & 2002 May 04 & 5393 \\
HD91824  & O5C095010 & 2000 Mar 23 & 360  & A1180802000 & 2000 Feb 06 & 4649 \\
HD91983  & O5C08N010 & 2000 May 18 & 1440 & B0710402000 & 2001 May 27 & 4308 \\
HD111934 & O5C03N010 & 2000 Mar 27 & 720  &     ---     &     ---     & ---  \\
HD116852 & O5C01C010 & 2000 Jun 25 & 360  & P1013801000 & 2000 May 27 & 7212 \\
HD122879 & O5C037010 & 2000 May 08 & 360  & B0710501000 & 2002 Mar 03 & 1532 \\
HD148594 & O5C04A010 & 2000 May 21 & 720  & P2310101000 & 2001 Aug 19 & 4016 \\
HD152590 & O5C08P010 & 2000 Mar 03 & 1440 & B0710602000 & 2001 Aug 12 & 2773 \\
         &    ---    &     ---     & ---  & B0710601000 & 2001 Jul 08 & 1389 \\
HD156110 & O5C01K010 & 1999 Nov 12 & 360  & B0710701000 & 2001 May 08 & 503  \\
HD157857 & O5C04D010 & 2000 Jun 03 & 720  & P1027501000 & 2000 Sep 02 & 4022 \\
HD165955 & O63599010 & 2001 May 11 & 1440 & P1027901000 & 2000 Aug 17 & 4140 \\
HD190918 & O6359J010 & 2000 Oct 12 & 720  & P1028501000 & 2000 Jul 18 & 5510 \\
HD192035 & O6359K010 & 2001 Feb 21 & 1440 & P1028603000 & 2000 Jun 21 & 6624 \\
         &    ---    &     ---     & ---  & P1028602000 & 2000 Jun 19 & 6148 \\
         &    ---    &     ---     & ---  & P1028601000 & 2000 Jun 17 & 4942 \\
HD192639 & O5C08T010 & 2000 Mar 01 & 1440 & P1162401000 & 2000 Jun 12 & 4834 \\
HD198478 & O5C06J010 & 2000 Apr 04 & 720  & P2350201000 & 2001 Jun 06 & 6628 \\
HD198781 & O5C049010 & 1999 Sep 06 & 360  & P2310201000 & 2001 Jul 21 & 844  \\
HD201345 & O5C050010 & 1999 Nov 12 & 360  & P1223001000 & 2000 Jun 13 & 5104 \\
HD203532 & O5C01S010 & 2000 Jun 29 & 720  & B0710801000 & 2001 Aug 23 & 4748 \\
HD206773 & O5C04T010 & 1999 Aug 12 & 360  & B0710901000 & 2001 Jul 19 & 4399 \\
HD208440 & O5C06M010 & 1999 Nov 05 & 720  & B0300401000 & 2001 Aug 03 & 9777 \\
HD210809 & O5C01V010 & 1999 Oct 30 & 720  & P1223103000 & 2000 Aug 08 &10065 \\
         & O6359T010 & 2001 Apr 24 & 720  & P1223102000 & 2000 Aug 07 & 7097 \\
         &    ---    &     ---     & ---  & P1223101000 & 2000 Aug 05 & 5508 \\
HD212791 & O5C04Q010 & 2000 Apr 25 & 1440 & Z9017701000 & 2002 Jun 30 & 2261 \\
HD220057 & O5C01X010 & 2000 Mar 24 & 360  & Z9017801000 & 2002 Aug 30 & 3037 \\
HD232522 & O5C08J010 & 1999 Oct 09 & 1440 & P1220101000 & 1999 Nov 28 &16800 \\
HD308813 & O63559010 & 2001 Feb 18 & 1440 & P1221903000 & 2000 Mar 27 & 5667 \\
         &    ---    &     ---     & ---  & P1221902000 & 2000 Mar 25 & 4340 \\
         &    ---    &     ---     & ---  & P1221901000 & 2000 Mar 23 & 4257 \\
BD $+$53 2820 &O6359Q010& 2001 Feb 05& 1440& M1141301000 & 2000 Aug 06 & 2529 \\
         &    ---    &     ---     & ---  & P1223203000 & 2000 Aug 08 & 5147 \\
         &    ---    &     ---     & ---  & P1223202000 & 2000 Aug 07 & 7119 \\
         &    ---    &     ---     & ---  & P1223201000 & 2000 Aug 06 & 5815 \\
CPD $-$69 1743&O63566010& 2001 Feb 20& 1440& P1013701000 & 2000 Apr 06 & 3567 \\
\enddata

%\tablecomments{The three oxygen column densities listed above refer to the
%values determined by apparent optical depth analysis (subscript $A$), profile
%fitting (subscript $P$), and the adopted result.}

\end{deluxetable}

\clearpage

\begin{deluxetable}{lllr@{$\;\!$}lr@{$\;\!$}lr@{$\;\!$}lr@{$\;\!$}l}
\rotate
\tablecaption{New Oxygen Measurements \label{oxytab}}
\tablewidth{0pt}
\tablehead{
&\colhead{Component Velocities}&\colhead{$b$-values}&
\multicolumn{2}{c}{$W_{\lambda1356}$}&\multicolumn{2}{c}{log$_{10}N$(O)$_A$}&
\multicolumn{2}{c}{log$_{10}N$(O)$_P$}&\multicolumn{2}{c}{log$_{10}N$(O)}\\
\colhead{Star}&\colhead{(km s$^{-1}$)}&\colhead{(km s$^{-1}$)}&
\multicolumn{2}{c}{(m\AA)}&\multicolumn{2}{c}{(cm$^{-2}$)}&
\multicolumn{2}{c}{(cm$^{-2}$)}&
\multicolumn{2}{c}{(cm$^{-2}$)}\\
}
\startdata
         &                                  &                         &
    &      &      &      &     &      &     &      \\
HD1383   &$-$62.5 $-$58.0 $-$50.5 $-$45.5 $-$40.6&2.2 2.7 2.4 2.6 2.8 &
24.1&(2.8) & 18.15&(0.05)&18.15&(0.03)&18.15&(0.06)\\
         &$-$36.1 $-$31.7 $-$25.7 $-$20.6 $-$15.6&2.7 1.3 2.2 2.3 2.4 &
    &      &      &      &     &      &     &      \\
         &$-$10.8  $-$6.0  $-$2.1                &2.3 2.4 2.3         &
    &      &      &      &     &      &     &      \\
HD12323  &$-$45.9 $-$33.8 $-$16.9  $-$8.7  $-$1.1&3.5 3.3 4.0 2.5 3.1 &
18.9&(1.6) & 18.03&(0.04)&18.02&(0.04)&18.02&(0.06)\\
HD13268  &$-$37.4 $-$19.1  $-$9.8  $-$2.2        &4.3 4.1 3.0 3.6     &
23.3&(2.6) & 18.12&(0.05)&18.13&(0.04)&18.13&(0.06)\\
HD14434  &$-$78.6 $-$63.8 $-$52.6 $-$40.8 $-$25.0&2.8 4.4 5.0 3.8 3.0 &
27.2&(3.6) & 18.19&(0.05)&18.18&(0.07)&18.18&(0.08)\\
         &$-$18.5  $-$9.2  $-$1.4                &3.2 3.7 3.3         &
    &      &      &      &     &      &     &      \\
HD36841  &   29.9                                &3.2                 &
 7.9&(1.1) & 17.65&(0.06)&17.63&(0.06)&17.63&(0.08)\\
HD37367  &   11.5    16.5                        &2.3 2.7             &
16.4&(1.1) & 18.03&(0.08)&18.06&(0.02)&18.06&(0.08)\\
HD43818  &    4.0     8.8    13.4    17.8    21.1&1.7 2.0 2.8 1.8 1.6 &
26.2&(1.4) & 18.22&(0.02)&18.22&(0.02)&18.22&(0.03)\\
         &   25.3                                &2.4                 &
    &      &      &      &     &      &     &      \\
HD52266  &    5.2    11.3    19.1    24.8    32.8&2.8 2.0 3.3 2.1 4.2 &
17.1&(1.3) & 18.00&(0.03)&18.01&(0.01)&18.01&(0.03)\\
         &   42.0                                &1.6                 &
    &      &      &      &     &      &     &      \\
HD63005  &   28.7    38.4                        &4.6 3.8             &
14.9&(0.7) & 17.94&(0.02)&17.95&(0.03)&17.95&(0.04)\\
HD71634  &    4.4     8.1    12.2    17.6    22.1&2.0 1.6 4.0 3.0 2.7 &
 5.4&(0.6) & 17.49&(0.05)&17.49&(0.03)&17.49&(0.06)\\
HD72754  &   17.1    20.8    26.3                &1.4 1.5 2.7         &
 8.7&(1.1) & 17.72&(0.05)&17.74&(0.04)&17.74&(0.06)\\
HD79186  &    6.8    11.8    19.0    24.8    30.2&1.2 2.6 2.5 1.5 1.4 &
18.0&(1.2) & 18.03&(0.03)&18.04&(0.02)&18.04&(0.04)\\
HD91824  &$-$18.7 $-$13.7  $-$7.0  $-$1.1     8.4&1.5 2.7 2.8 2.0 3.3 &
17.3&(1.5) & 18.00&(0.04)&18.00&(0.03)&18.00&(0.05)\\
         &   15.4                                &1.2                 &
    &      &      &      &     &      &     &      \\
HD91983  &$-$16.3 $-$11.2  $-$3.7     7.6    11.9&2.2 1.4 2.8 2.9 2.0 &
15.6&(1.6) & 17.95&(0.04)&17.96&(0.02)&17.96&(0.04)\\
         &   18.4                                &2.0                 &
    &      &      &      &     &      &     &      \\
HD111934 &$-$34.8 $-$24.2 $-$19.5 $-$14.6  $-$9.1&1.9 2.5 2.9 2.2 2.3 &
25.4&(2.8) & 18.17&(0.05)&18.18&(0.07)&18.17&(0.08)\\
         & $-$3.9     3.3     7.4    15.0        &1.4 2.4 2.4 3.1     &
    &      &      &      &     &      &     &      \\
HD116852 &    3.4     9.7    15.4                &2.1 1.8 1.2         &
 6.6&(0.6) & 17.57&(0.04)&17.57&(0.03)&17.57&(0.05)\\
HD122879 &$-$28.4 $-$24.1 $-$19.6 $-$15.6  $-$9.9&1.3 2.7 2.2 2.3 2.2 &
23.8&(1.7) & 18.14&(0.03)&18.14&(0.05)&18.14&(0.06)\\
         & $-$3.8  $-$0.0     4.3     9.4        &2.0 2.1 1.8 1.2     &
    &      &      &      &     &      &     &      \\
HD148594 &$-$10.5  $-$4.7                        &1.6 2.2             &
10.5&(0.6) & 17.82&(0.02)&17.86&(0.02)&17.86&(0.03)\\
HD156110 &$-$25.3 $-$17.9                        &1.8 2.3             &
 3.4&(0.6) & 17.26&(0.07)&17.25&(0.05)&17.25&(0.09)\\
HD157857 &$-$19.9 $-$12.9  $-$7.3                &1.0 2.5 3.7         &
19.9&(0.8) & 18.10&(0.02)&18.11&(0.03)&18.11&(0.04)\\
HD165955 &$-$16.2  $-$3.1    11.2                &2.9 4.3 2.9         &
11.1&(1.8) & 17.78&(0.07)&17.79&(0.06)&17.79&(0.09)\\
HD190918 &$-$24.2 $-$13.3     1.2    14.1        &4.5 5.7 4.0 5.5     &
23.4&(2.2) & 18.12&(0.04)&18.12&(0.07)&18.12&(0.08)\\
HD192035 &$-$22.5 $-$12.5  $-$3.5                &5.1 4.9 3.8         &
17.7&(1.3) & 18.01&(0.03)&18.01&(0.03)&18.01&(0.04)\\
HD192639 &$-$16.1  $-$9.5                        &2.7 2.5             &
21.0&(1.0) & 18.13&(0.02)&18.14&(0.03)&18.14&(0.04)\\
HD198478 &$-$21.6 $-$17.3 $-$12.1  $-$7.4        &2.2 2.2 2.0 1.8     &
16.5&(1.0) & 18.00&(0.03)&18.02&(0.04)&18.02&(0.05)\\
HD198781 &$-$46.5 $-$41.1 $-$35.8 $-$30.9 $-$27.5&2.0 1.5 2.1 1.5 1.0 &
 9.7&(0.8) & 17.75&(0.04)&17.77&(0.05)&17.77&(0.06)\\
         &$-$21.7 $-$15.8  $-$9.2  $-$2.7        &3.7 2.9 2.3 2.8     &
    &      &      &      &     &      &     &      \\
HD201345 &$-$25.3 $-$21.1 $-$16.8 $-$13.1  $-$8.6&1.3 1.2 2.3 2.0 2.5 &
 8.2&(1.2) & 17.66&(0.05)&17.66&(0.04)&17.66&(0.06)\\
         & $-$4.3                                &2.7                 &
    &      &      &      &     &      &     &      \\
HD206773 &$-$26.3 $-$20.6 $-$15.2 $-$11.0  $-$6.0&2.1 2.4 1.9 1.8 3.1 &
13.0&(1.3) & 17.88&(0.04)&17.90&(0.04)&17.90&(0.05)\\
HD208440 &$-$29.1 $-$23.1 $-$17.8 $-$12.8  $-$6.3&1.3 2.5 1.7 2.8 1.4 &
15.9&(1.5) & 17.96&(0.04)&17.96&(0.06)&17.96&(0.07)\\
HD210809 &$-$52.1 $-$46.1 $-$38.5 $-$34.2 $-$28.3&2.2 1.6 2.1 0.9 1.0 &
18.6&(2.7) & 18.02&(0.06)&17.99&(0.04)&17.99&(0.07)\\
         &$-$20.1 $-$13.8  $-$9.1  $-$1.5        &2.4 2.0 2.5 1.5     &
    &      &      &      &     &      &     &      \\
HD212791 &$-$27.9 $-$24.9 $-$21.0 $-$16.3 $-$11.0&2.8 2.2 2.6 2.6 2.4 &
12.2&(1.4) & 17.83&(0.05)&17.83&(0.03)&17.83&(0.06)\\
         &$-$6.8                                 &3.5                 &
    &      &      &      &     &      &     &      \\
HD220057 &$-$16.7 $-$10.6  $-$6.3                &2.6 3.0 3.0         &
11.2&(1.0) & 17.82&(0.04)&17.79&(0.04)&17.79&(0.06)\\
HD232522 &$-$22.5 $-$16.8  $-$9.6  $-$6.1        &2.0 2.0 2.2 2.2     &
10.2&(1.2) & 17.76&(0.05)&17.78&(0.07)&17.78&(0.08)\\
HD308813 &$-$31.9 $-$16.2  $-$5.6     4.9    19.2&4.7 4.1 3.7 5.0 2.5 &
15.0&(1.9) & 17.92&(0.05)&17.90&(0.06)&17.90&(0.08)\\
BD+53 2820&$-$58.8 $-$44.3 $-$34.6 $-$28.4 $-$16.3&3.1 4.1 3.4 2.3 4.9 &
19.8&(3.9) & 18.05&(0.08)&18.05&(0.09)&18.05&(0.12)\\
         & $-$6.8 6.1 21.2                    &3.6 4.1                &
    &      &      &      &     &      &     &      \\
CPD-69 1743&$-$23.7 $-$13.5 $-$2.5    7.4    20.2&2.6 3.4 3.9 5.0 1.8 &
11.5&(2.7) & 17.80&(0.09)&17.80&(0.09)&17.80&(0.12)\\
\enddata

\tablecomments{The three oxygen column densities listed above refer to the
values determined by apparent optical depth analysis (subscript $A$), profile
fitting (subscript $P$), and the adopted result. Table entries are restricted
to sight lines for which we have not previously published oxygen profile data.}

\end{deluxetable}

\clearpage

\begin{deluxetable}{lr@{$\;\!$}lr@{$\;\!$}lr@{$\;\!$}lr@{$\;\!$}lr@{$\;\!$}lr@{.}lc}
%\rotate
%\tabletypesize{\scriptsize}
\tabletypesize{\footnotesize}
\tablecaption{Oxygen and Hydrogen Sight Line Properties \label{oxyhydtab}}
\tablewidth{0pt}
\tablehead{
&\multicolumn{2}{c}{\hspace*{-0.15cm}log$_{10}$[$N$(\ion{H}{1})]}&
\multicolumn{2}{c}{\hspace*{-0.1cm}log$_{10}$[$N$(H$_2$)]}&
\multicolumn{2}{c}{log$_{10}$[$N$(H)]}&
\multicolumn{2}{c}{\hspace*{-0.05cm}log$_{10}$[$N$(O)]}&
\multicolumn{2}{c}{ }&
\multicolumn{2}{c}{\hspace*{-0.2cm}log$_{10}{\langle}n_{\rm H}\rangle$}&
\colhead{ }\\
\colhead{Star}&
\multicolumn{2}{c}{(cm$^{-2}$)}&
\multicolumn{2}{c}{(cm$^{-2}$)}&
\multicolumn{2}{c}{(cm$^{-2}$)}&
\multicolumn{2}{c}{(cm$^{-2}$)}&
\multicolumn{2}{c}{log$_{10}$[O/H]}&
\multicolumn{2}{c}{(cm$^{-3}$)}&
\colhead{log$_{10}f$(H$_2$)}\\
}
\startdata
HD1383         & 21.42&(0.09) & 20.45&(0.07) & 21.50&(0.08) &
18.15&(0.06) & $-$3.35&(0.10) & $-$0&45 & $-$0.75\\
HD12323        & 21.18&(0.09) & 20.32&(0.08) & 21.29&(0.07) &
18.02&(0.06) & $-$3.27&(0.09) & $-$0&56 & $-$0.67\\
HD13268        & 21.32&(0.09) & 20.42&(0.07) & 21.42&(0.07) &
18.13&(0.06) & $-$3.29&(0.09) & $-$0&43 & $-$0.70\\
HD14434        & 21.37&(0.09) & 20.47&(0.07) & 21.47&(0.07) &
18.18&(0.08) & $-$3.29&(0.10) & $-$0&38 & $-$0.70\\
HD27778        & 21.10&(0.12) & 20.72&(0.08) & 21.36&(0.08) &
17.83&(0.04) & $-$3.53&(0.09) &    0&53 & $-$0.34\\
%HD36841\tablenotemark{a}& 21.69&(0.15) & \multicolumn{2}{c}{...} &
%21.23&(0.18) & 17.63&(0.08) & $-$3.60&(0.20) & 0&18 & ... \\
HD37021        & 21.68&(0.12) & \multicolumn{2}{c}{...} &
21.68&(0.12) & 18.09&(0.03) & $-$3.59&(0.12) & 0&61 & ... \\
HD37061        & 21.73&(0.09) & \multicolumn{2}{c}{...} &
21.73&(0.09) & 18.23&(0.02) & $-$3.50&(0.09) & 0&66 & ... \\
HD37367        & 21.28&(0.09) & 20.53&(0.09) & 21.41&(0.07) &
18.06&(0.08) & $-$3.35&(0.10) &    0&36 & $-$0.58\\
HD37903        & 21.16&(0.09) & 20.85&(0.07) & 21.46&(0.06) &
17.88&(0.02) & $-$3.58&(0.06) &    0&41 & $-$0.31\\
HD43818\tablenotemark{a}& 21.60&(0.12) & \multicolumn{2}{c}{...} &
21.71&(0.11) & 18.22&(0.03) & $-$3.49&(0.11) & 0&04 & ... \\
HD63005        & 21.24&(0.06) & 20.23&(0.09) & 21.32&(0.05) &
17.95&(0.04) & $-$3.37&(0.06) & $-$0&57 & $-$0.79\\
HD72754        & 21.18&(0.12) & 20.35&(0.10) & 21.29&(0.10) &
17.74&(0.06) & $-$3.55&(0.11) & $-$0&04 & $-$0.64\\
HD75309        & 21.08&(0.09) & 20.20&(0.12) & 21.18&(0.08) &
17.73&(0.05) & $-$3.45&(0.09) & $-$0&55 & $-$0.68\\
HD79186        & 21.20&(0.09) & 20.72&(0.09) & 21.42&(0.07) &
18.04&(0.04) & $-$3.38&(0.08) & $-$0&31 & $-$0.40\\
HD91824        & 21.12&(0.06) & 19.85&(0.07) & 21.16&(0.05) &
18.00&(0.05) & $-$3.16&(0.07) & $-$0&73 & $-$1.01\\
HD91983        & 21.17&(0.09) & 20.14&(0.07) & 21.24&(0.08) &
17.96&(0.04) & $-$3.28&(0.09) & $-$0&65 & $-$0.80\\
HD116852       & 20.96&(0.09) & 19.79&(0.11) & 21.02&(0.08) &
17.57&(0.05) & $-$3.45&(0.09) & $-$1&15 & $-$0.93\\
HD122879       & 21.26&(0.12) & 20.24&(0.09) & 21.34&(0.10) &
18.14&(0.06) & $-$3.20&(0.11) & $-$0&47 & $-$0.80\\
HD147888\tablenotemark{b}& 21.71&(0.09) & 20.57&(0.15) & 21.77&(0.08) &
18.18&(0.02) & $-$3.59&(0.08) &    1&11 & $-$0.90\\
HD148594\tablenotemark{a}& 21.80&(0.15) & 19.88&(0.07) & 21.39&(0.10) &
17.86&(0.03) & $-$3.53&(0.10) &    0&72 & $-$1.21\\
HD152590       & 21.37&(0.06) & 20.47&(0.07) & 21.47&(0.05) &
18.06&(0.05) & $-$3.41&(0.07) & $-$0&28 & $-$0.70\\
HD157857       & 21.26&(0.09) & 20.68&(0.10) & 21.44&(0.07) &
18.11&(0.04) & $-$3.33&(0.08) & $-$0&41 & $-$0.46\\
HD165955       & 21.11&(0.06) & 16.53&(0.04) & 21.11&(0.06) &
17.79&(0.09) & $-$3.32&(0.11) & $-$0&59 & $-$4.28\\
HD175360\tablenotemark{a}& 21.53&(0.09) & \multicolumn{2}{c}{...} &
21.02&(0.11) & 17.59&(0.06) & $-$3.43&(0.12) & 0&11 & ... \\
HD185418       & 21.19&(0.09) & 20.71&(0.12) & 21.41&(0.07) &
18.06&(0.05) & $-$3.35&(0.08) &    0&08 & $-$0.40\\
HD190918       & 21.38&(0.06) & 19.84&(0.08) & 21.40&(0.06) &
18.12&(0.08) & $-$3.28&(0.10) & $-$0&45 & $-$1.26\\
HD192035       & 21.20&(0.09) & 20.62&(0.07) & 21.38&(0.07) &
18.01&(0.04) & $-$3.37&(0.08) & $-$0&50 & $-$0.46\\
HD192639       & 21.29&(0.09) & 20.73&(0.10) & 21.48&(0.07) &
18.14&(0.04) & $-$3.34&(0.08) & $-$0&27 & $-$0.45\\
HD198478       & 21.32&(0.15) & 20.87&(0.15) & 21.55&(0.11) &
18.02&(0.05) & $-$3.53&(0.12) &    0&16 & $-$0.38\\
HD198781       & 20.91&(0.09) & 20.48&(0.07) & 21.15&(0.06) &
17.77&(0.06) & $-$3.38&(0.08) & $-$0&13 & $-$0.37\\
HD201345       & 20.97&(0.09) & 19.55&(0.13) & 21.00&(0.08) &
17.66&(0.06) & $-$3.34&(0.10) & $-$0&90 & $-$1.15\\
HD203532       & 21.27&(0.09) & 20.64&(0.08) & 21.44&(0.07) &
17.85&(0.02) & $-$3.59&(0.07) &    0&55 & $-$0.50\\
HD206773       & 21.09&(0.06) & 20.44&(0.10) & 21.25&(0.05) &
17.90&(0.05) & $-$3.35&(0.07) & $-$0&03 & $-$0.51\\
HD207198       & 21.53&(0.07) & 20.83&(0.10) & 21.68&(0.09) &
18.13&(0.04) & $-$3.55&(0.10) &    0&40 & $-$0.55\\
HD208440       & 21.23&(0.09) & 20.29&(0.07) & 21.32&(0.08) &
17.96&(0.07) & $-$3.36&(0.10) &    0&04 & $-$0.73\\
HD210809       & 21.29&(0.09) & 20.00&(0.09) & 21.33&(0.08) &
17.99&(0.07) & $-$3.34&(0.10) & $-$0&70 & $-$1.03\\
HD212791       & 21.21&(0.09) & 19.42&(0.11) & 21.22&(0.09) &
17.83&(0.06) & $-$3.39&(0.11) &    0&16 & $-$1.50\\
HD220057       & 21.17&(0.09) & 20.28&(0.07) & 21.27&(0.07) &
17.79&(0.06) & $-$3.48&(0.09) &    0&14 & $-$0.69\\
HD232522       & 21.08&(0.06) & 20.22&(0.09) & 21.19&(0.05) &
17.78&(0.08) & $-$3.41&(0.09) & $-$0&92 & $-$0.67\\
HD308813       & 21.20&(0.09) & 20.25&(0.07) & 21.29&(0.08) &
17.90&(0.08) & $-$3.39&(0.11) & $-$0&18 & $-$0.74\\
BD $+$53 2820  & 21.35&(0.09) & 20.01&(0.11) & 21.39&(0.08) &
18.05&(0.12) & $-$3.34&(0.12) & $-$0&64 & $-$1.08\\
CPD $-$69 1743 & 21.11&(0.09) & 19.90&(0.09) & 21.16&(0.08) &
17.80&(0.12) & $-$3.36&(0.11) & $-$1&15 & $-$0.96\\
$\gamma$ Cas   & 20.17&(0.06) & 17.51&\hspace*{0.3cm}...&
20.18&(0.05) & 16.76&(0.04) & $-$3.42&(0.06) & $-$0&58 & $-$2.37\\
$\zeta$ Per    & 20.81&(0.04) & 20.67&(0.10) & 21.20&(0.06) &
17.71&(0.05) & $-$3.49&(0.07) &    0&11 & $-$0.23\\
$\epsilon$ Per & 20.42&(0.06) & 19.53&(0.15) & 20.52&(0.06) &
17.03&(0.04) & $-$3.49&(0.07) & $-$0&46 & $-$0.68\\
$\xi$ Per      & 21.08&(0.06) & 20.53&(0.08) & 21.27&(0.05) &
17.81&(0.05) & $-$3.46&(0.07) &    0&18 & $-$0.44\\
23 Ori         & 20.74&(0.08) & 18.30&(0.11) & 20.74&(0.08) &
17.32&(0.10) & $-$3.42&(0.12) & $-$0&22 & $-$2.14\\
$\delta$ OriA  & 20.20&(0.05) & 14.58&\hspace*{0.3cm}...&
20.20&(0.05) & 16.67&(0.05) & $-$3.53&(0.07) & $-$0&86 & $-$5.22\\
$\lambda$ Ori  & 20.79&(0.08) & 19.11&(0.11) & 20.81&(0.07) &
17.33&(0.06) & $-$3.48&(0.09) & $-$0&38 & $-$1.40\\
$\iota$ Ori    & 20.16&(0.05) & 14.69&\hspace*{0.3cm}...&
20.16&(0.05) & 16.76&(0.07) & $-$3.40&(0.08) & $-$1&03 & $-$5.17\\
$\epsilon$ Ori & 20.46&(0.07) & 16.57&\hspace*{0.3cm}...&
20.46&(0.07) & 16.98&(0.05) & $-$3.48&(0.08) & $-$0&73 & $-$3.59\\
$\kappa$ Ori   & 20.53&(0.04) & 15.68&\hspace*{0.3cm}...&
20.53&(0.04) & 17.03&(0.04) & $-$3.50&(0.06) & $-$0&66 & $-$4.55\\
15 Mon         & 20.36&(0.06) & 15.55&\hspace*{0.3cm}...&
20.36&(0.06) & 16.84&(0.09) & $-$3.52&(0.11) & $-$0&98 & $-$4.51\\
$\tau$ CMa     & 20.71&(0.02) & 15.48&\hspace*{0.3cm}...&
20.71&(0.04) & 17.33&(0.04) & $-$3.39&(0.05) & $-$0&96 & $-$4.93\\
$\zeta$ Oph    & 20.71&(0.02) & 20.65&(0.05) & 21.15&(0.03) &
17.63&(0.04) & $-$3.52&(0.05) &    0&53 & $-$0.20\\
$\gamma$ Ara   & 20.71&(0.06) & 19.24&(0.13) & 20.74&(0.06) &
17.33&(0.04) & $-$3.41&(0.07) & $-$0&58 & $-$1.20\\
\enddata

\tablenotetext{a}{Atomic hydrogen column densities cannot reliably be measured
toward HD148594 and HD175360 because of significant stellar contamination of
the interstellar \ion{H}{1} absorption features. The \ion{H}{1} abundances
listed above for these paths refer to the combined stellar and interstellar
profiles in each spectrum. \textit{FUSE} data are not available for HD43818;
consequently, a measurement of $N$(H$_@$) cannot be made for this sight line.
The total hydrogen column density for each of these paths was thus derived from
krypton measurements assuming the constancy of the Kr/H abundance ratio within
several hundred parsecs \citep{car03}. The krypton abundance measurement for
HD43818 is somewhat poorer in quality (log$N$(Kr) = 12.69$\pm$0.09) than those
for HD148594 and HD175360 \citep{car01,car03}, necessitating larger $N$(H)
error bars for this sight line in the Figures and a correspondingly lower
weight in the Boltzmann function fitting procedure.}
\tablenotetext{b}{As stated in the text, HD147933 was used as a proxy for
hydrogen abundances toward HD147888. Of note, the atomic component determined
from Ly$\alpha$ absorption in the STIS UV spectrum for HD147888 matched the
value for HD147933 determined by a weighted mean of abundances determined by
\citet{boh78} and \citet{dip94}.}

\tablecomments{Only sight lines with reliable measurements of both oxygen and
hydrogen absorption or those for which a reasonable proxy for $N$(H) was
available have been included in this table; sight lines identified by their HD
numbers or BD or CPD designations were observed using STIS--the remainder were
observed with GHRS.}

\end{deluxetable}

\begin{deluxetable}{llcr@{.}lr@{.}lc}
\tablecaption{Summary of Sight Line Properties \label{apptab}}
\tablewidth{0pt}
\tablehead{
Sight Line&Alternate&Spectral&\multicolumn{2}{c}{}&\multicolumn{2}{c}{Distance}&
Group\\
(HD)&Name&Type&\multicolumn{2}{c}{$E(\bv)$}&\multicolumn{2}{c}{(kpc)}&
Membership\\
}
\startdata
1383   &BD$+$60 25      &\ion{B1}{2}     &0&51&2&86&Field   \\
12323  &BD$+$54 441     &\ion{O9}{5}     &0&41&2&29&Per OB1 \\
13268  &BD$+$55 534     &\ion{O8}{5}     &0&44&2&29&Per OB1 \\
14434  &BD$+$56 567     &\ion{O5.5}{5}nfp&0&48&2&29&Per OB1 \\
27778  &62 Tau          &\ion{B3}{5}     &0&38&0&22&Gould B.\\
36841  &BD$-$00 1002    &O8              &0&34&0&36&Ori OB1b\\
37021  &$\theta^1$ Ori B&\ion{B0}{5}     &0&44&0&38&Ori OB1d\\
37061  &$\nu$ Ori       &\ion{B1}{5}     &0&54&0&38&Ori OB1d\\
37367  &BD$+$29 947     &\ion{B2}{4-}V   &0&40&0&36&Field   \\
37903  &BD$-$02 1345    &\ion{B1.5}{5}   &0&35&0&36&Ori OB1b\\
43818  &LU Gem          &\ion{B0}{2}     &0&58&1&51&Gem OB1 \\
52266  &BD$-$05 1912    &\ion{O9}{5}     &0&30&1&73&Field   \\
63005  &CPD$-$26 2525   &\ion{O6}{5}f    &0&30&2&51&Pup OB1 \\
71634  &CPD$-$57 1513   &\ion{B7}{4}     &0&10&0&40&Unknown \\
72754  &FY Vel&\ion{B2}{1}a{\thinspace}pe&0&36&0&69&Unknown \\
75309  &CPD$-$45 3056   &\ion{B1}{2}p    &0&27&1&75&Vel OB1 \\
79186  &CPD$-$44 5206   &\ion{B5}{1}a    &0&30&1&75&Vel OB1 \\
91824  &CPD$-$57 3463   &O7              &0&27&2&51&Car OB1 \\
91983  &CPD$-$57 3516   &\ion{O9.5/B0}{1}b&0&26&2&51&Car OB1\\
111934 &CPD$-$59 4543   &\ion{B3}{1}b    &0&32&2&40&Cen OB1 \\
116852 &CPD$-$78 813    &\ion{O9}{3}     &0&22&4&80&Runaway \\
122879 &CPD$-$59 5395   &\ion{B0}{1}a    &0&36&2&10&Field   \\
147888 &$\rho$ Oph D    &\ion{B3}{5}     &0&52&0&15&Sco OB2-2\\
148594 &CPD$-$27 5408   &\ion{B8}{5}     &0&21&0&15&Sco OB2-2\\
152590 &CPD$-$40 7624   &\ion{O7.5}{5}   &0&38&1&80&Sco OB1 \\
156110 &BD$+$45 2509    &\ion{B3}{5}n    &0&03&0&72&Unknown \\
157857 &BD$-$10 4493    &\ion{O6.5}{3}f  &0&50&2&30&Runaway \\
165955 &CPD$-$34 7625   &\ion{B3}{5}n    &0&14&1&64&Field   \\
175360 &CPD$-$23 7307   &\ion{B6}{3}     &0&12&0&27&Unknown \\
185418 &BD$+$16 3928    &\ion{B0.5}{5}   &0&51&0&69&Field   \\
190918 &BD$+$35 3953    &\ion{O9.5}{1}ab &0&41&2&29&Cyg OB3 \\
192035 &BD$+$47 3038    &\ion{B0}{4}     &0&35&2&44&Field   \\
192639 &BD$+$36 3958    &O8e             &0&66&1&82&Cyg OB1 \\
198478 &55 Cyg          &\ion{B3}{1}a    &0&53&0&79&Cyg OB7 \\
198781 &BD$+$63 1663    &\ion{B0.5}{5}   &0&35&0&62&Cep OB2 \\
201345 &BD$+$32 4060    &\ion{O9}{5}     &0&18&2&57&Field   \\
203532 &CPD$-$83 716    &\ion{B3}{4}     &0&33&0&25&Unknown \\
206773 &BD$+$57 2374    &\ion{B0}{5}pe   &0&44&0&62&Cep OB2 \\
207198 &BD$+$61 2193    &\ion{O9}{2}     &0&62&0&62&Cep OB2 \\
208440 &BD$+$61 2217    &\ion{B1}{5}     &0&34&0&62&Cep OB2 \\
210809 &BD$+$51 3281    &\ion{O9}{1}ab   &0&33&3&47&Cep OB1 \\
212791 &BD$+$51 3372    &B3ne\tablenotemark{a}&0&05&0&37&Unknown \\
220057 &BD$+$60 2521    &\ion{B3}{4}     &0&27&0&44&NGC 7654\\
232522 &BD$+$54 372     &\ion{B1}{2}     &0&18&4&19&Field   \\
308813 &CPD $-$62 2158  &\ion{O9.5}{5}   &0&30&0&96&IC 2944 \\
       &BD $+$53 2820   &\ion{B0}{4}n    &0&36&3&47&Cep OB1 \\
       &CPD $-$69 1743  &\ion{B0.5}{3}n  &0&30&6&56&Unknown \\
\enddata

\tablenotetext{a}{\citet{ham91,koh99}.}

\end{deluxetable}

\end{document}